\documentclass[fleqn,usenatbib]{mnras}

\usepackage{mathptmx}

\usepackage[T1]{fontenc}

\DeclareRobustCommand{\VAN}[3]{#2}
\let\VANthebibliography\thebibliography
\def\thebibliography{\DeclareRobustCommand{\VAN}[3]{##3}\VANthebibliography}

\usepackage{graphicx}	
\usepackage{amsmath}	
\usepackage{amssymb}	
\graphicspath{{./img/}}  
\usepackage{bm}
\usepackage{appendix}
\usepackage{subfigure}
\usepackage{float}
\usepackage{supertabular}
\usepackage{geometry}
\usepackage{color}

\newcommand{\kmps}{$\text{km s}^{-1}$}

\title[Bimodal orientation distribution and head-tail asymmetry of the filaments]{Bimodal orientation distribution and head-tail asymmetry of a sample of filamentary molecular clouds}

\author[Wen Ge et al.]{
Wen Ge,$^{1,2}$
Fujun Du,$^{1,2}$\thanks{E-mail: fjdu@pmo.ac.cn}
Lixia Yuan$^{1}$
\\
$^{1}$Purple Mountain Observatory and Key Laboratory of Radio Astronomy, Chinese Academy of Sciences, 10 Yuanhua Road, Qixia District, \\Nanjing 210023, People's Republic of China\\
$^{2}$School of Astronomy and Space Science, University of Science and Technology of China, Hefei 230026, People's Republic of China\\
}

\date{Accepted 2024 February 27. Received 2024 January 22; in original form 2023 September 14}

\pubyear{2024}

\begin{document}
\label{firstpage}
\pagerange{\pageref{firstpage}--\pageref{lastpage}}
\maketitle

\begin{abstract}
The morphology of molecular clouds is crucial for understanding their origin and evolution. In this work, we investigate the morphology of the filamentary molecular clouds (filaments for short) using a portion of the $^{12}\text{CO} (J=1-0)$ data from the Milky Way Imaging Scroll Painting (MWISP) project. The data cover an area spanning $104.75^\circ <l< 150.25^\circ , \vert b\vert < 5.25^\circ$ in Galactic coordinates, with $V_\text{LSR}$ ranging from $-95$ to 25 \kmps. Our primary focus is on the orientation and morphological asymmetry of the filaments. To achieve this, we apply several criteria on the data to create a sample of filaments with well-defined straight shape, and we use elliptical fitting to obtain the orientation of each filament, with an estimated error of $\sim1.6^\circ$ for the orientation. We find that the filament orientation with respect to the Galactic plane exhibits a bimodal distribution, a double-Gaussian fitting of which has two centres located at $-38.1^\circ $ and $42.0^\circ $, with 1$\sigma$ of the two Gaussian functions being 35.4$^\circ$ and 27.4$^\circ$. We do not find significant correlation between the orientation and other parameters, including the Galactic coordinates, radial velocity, velocity width, and physical scale. A considerable fraction of filaments ($\gtrsim$ 40 per cent) display head-tail asymmetry, which suggests that mass concentration tends to occur at one end of the filaments.
\end{abstract}




\section{Introduction}
Recent studies have demonstrated that molecular clouds display a variety of structural forms, one of the most prominent being filamentary molecular clouds \citep{2011A&A...529L...6A,2013A&A...550A..38P,2015A&A...584A..91K}. The results from the Gould Belt Survey with \textit{Herschel} have revealed numerous filamentary structures in Galactic molecular clouds, and a connection between the filamentary structure and the formation of dense cloud cores has been made \citep{2010A&A...518L...1P,2010ApJ...719L.185J,2010A&A...518L.100M,2010A&A...518L.102A}.

Filaments have been observed at different scales from $\lesssim$1 pc \citep{2013A&A...554A..55H,2018A&A...610A..77H} to hundreds of pc \citep{2013A&A...559A..34L,2014A&A...568A..73R,2022arXiv220503935P}.  Various investigations using diverse tracers have significantly enhanced our understanding of filaments, such as extinction \citep{1999A&A...345..965C,2009ApJ...700.1609M},  dust emission \citep{1994ApJ...423L..59A}, H\,{\textsc i} \citep{2006ApJ...652.1339M}, and CO emission from diffuse molecular gas \citep{2001ApJ...555..178F,2009A&A...500L..29H}. For instance, \cite{2016ApJS..226....9W} present a study with the dense gas tracers HCO$^+$ (3$-$2) and N$_2$H$^+$ (3$-$2), which includes a comprehensive catalog and physical parameters of 54 large-scale filaments. They find that the filaments are widely distributed across the Galactic disk, and most of the filaments are associated with major spiral arms.  \cite{2021ApJS..257...51Y} use $^{12}$CO as the tracer. They classified the morphology of 18190 molecular clouds visually.  Despite the lower fraction of filaments versus nonfilaments ($\sim $11 per cent \textit{vs} $\sim $89 per cent), filaments contribute $\sim $90 per cent to the total flux of the whole sample. In addition, filaments tend to have larger spatial scales than nonfilaments.

There are several possible explanations for the formation of filaments \citep{2022arXiv220309562H}, necessarily including some anisotropic factors such as large-scale gas motions sweeping up material \citep{2004ApJ...616..288B}, shock interactions induced by turbulence \citep{2016MNRAS.457..375F}, magnetic fields \citep{2013ApJ...774..128S}, stellar feedback \citep{2012A&A...541A..63P}, or Galactic shear \citep{2017MNRAS.470.4261D}. The plurality of these physical processes may explain why filaments are ubiquitous in ISM.

Morphological analysis is crucial for understanding the formation of filamentary molecular clouds, because the shaping of them by large-scale processes of a galaxy may be reflected in their morphological characteristics.
Some studies found that orientations of filaments tend to be either parallel or perpendicular to the directions of the local magnetic field \citep{2013MNRAS.436.3707L,2020MNRAS.496.4546H,2016A&A...586A.135P}, while other studies also showed that the magnetic field is usually aligned in parallel with the filaments in the molecular clouds \citep{2006ApJ...652.1339M,2014ApJ...789...82C}. The numerical simulation of \cite{2021MNRAS.503.5425B} showed a tendency for magnetic fields to be aligned perpendicularly to dense filaments, and parallel to low density filaments, while \cite{2021MNRAS.502.2285D} found in their simulations that filaments are perpendicular to the magnetic field in sub-Alfv{\'e}nic systems, whereas the opposite is true in super-Alfv{\'e}nic models.
Besides, the cold interstellar medium is observed to be organized in bubbles and filamentary structures \citep{2022arXiv220503935P}, and some investigations also indicate the expanding nature of these bubbles shapes the surrounding medium and possibly plays a role in the formation and evolution of interstellar filaments \citep{2011ApJ...731...13N,2015A&A...580A..49I}.
\cite{2013A&A...550A..38P} study the orientation of substructures within the dense, star-forming filament B211 in the Taurus molecular cloud. They find that the low-density striations are almost orthogonal to the B211 main filament, and very close to the local direction of the magnetic field.
\cite{2021A&A...651L...4S} perform a statistical study of the filamentary structure orientation in the CO emission data from the MWISP project.  They find that while most of the filamentary structures in CO do not show a preference to be parallel or perpendicular to the Galactic plane, a subset of them are indeed aligned with the Galactic plane.

In this work, we study the orientation and the head-tail asymmetry of filamentary molecular clouds. In Section \ref{Data and Sample selection}, we introduce the data we use, and we describe the procedures we follow to construct a sample of filaments with well-defined straight morphology.  In Section \ref{Bimodal orientation distribution of the sample}, we use the direction of the semi-major axis obtained from the elliptical fitting as the orientation of the filament, then we study the distribution of the orientation. In Section \ref{Head-tail asymmetry}, we study the integrated intensity along the major axis of the filaments. In addition to skewness, we also define a new parameter $p$ to measure the head-tail asymmetry of the filaments. We present a summary of our findings in Section \ref{Conclusion}.

\section{Data and sample selection} \label{Data and Sample selection}
The MWISP project is a systematic survey of CO and its isotopologues on the Galactic plane with the 13.7 m single-dish telescope of the Purple Mountain Observatory. Three molecular spectral lines of $^{12}$CO / $^{13}$CO / C$^{18}$O $J=1-0$ are observed at the same time. The typical system temperatures are $\sim$250 K for $^{12}$CO and $\sim$140 K for $^{13}$CO and C$^{18}$O, respectively \citep{2018AcASn..59....3S,2019ApJS..240....9S}. The data used in this work are in the form of a fits cube, where the first two dimensions represent the Galactic longitude and latitude, while the third dimension represents the local standard-of-rest velocity ($V\rm _{LSR}$). The spatial resolution of the data is $50^{\prime \prime}$ with an approximately Nyquist-sampling pixel size of $30^{\prime \prime}$, and the velocity resolution is 0.167 \kmps.

In this work, the 18190 molecular clouds that we use are obtained from the $^{12}\text{CO}(1-0)$ spectral line data using the DBSCAN algorithm \citep{1996kddm.conf..226E}, with Galactic longitudes  $104.75^\circ \le l \le150.25^\circ$, and Galactic latitudes $\vert b\vert$ $\leqslant$  $5.25^\circ$ \citep{2021ApJS..257...51Y}. DBSCAN has two parameters ($\epsilon $ and MinPts), where $\epsilon $ defines the neighborhood of voxels and MinPts defines core points, which are voxels whose numbers of neighboring voxels (including itself) are at least equal to MinPts. See \cite{2021A&A...645A.129Y} for details about the parameters of the DBSCAN algorithm. The minimum cutoff on the PPV data cubes is set to 2$\sigma $ ($\sim $1 K).

To study the orientation of filaments, we construct a sample with a simple straight morphology from the 18190 clouds of \cite{2021ApJS..257...51Y}, since only for such filaments a meaningful orientation can be defined. We first exclude those that have too large or too small angular scales; the former could be molecular cloud complexes, while the latter are limited by the angular resolution, thus an orientation is not well-defined for them. We also exclude the clouds with narrow velocity widths. If the velocity width is too narrow, the S/N of the image may not be high enough. Thus, we set $7 \times  7 \times 5$ ($3.5^{\prime} \times 3.5^{\prime} \times $ 0.835 \kmps) and $80 \times 80 \times 40$ ($40^{\prime} \times 40^{\prime} \times $ 6.680 \kmps) as the minimum and maximum cloud sizes in voxel, respectively.

\begin{figure}
    \centering
    \includegraphics[width=\columnwidth]{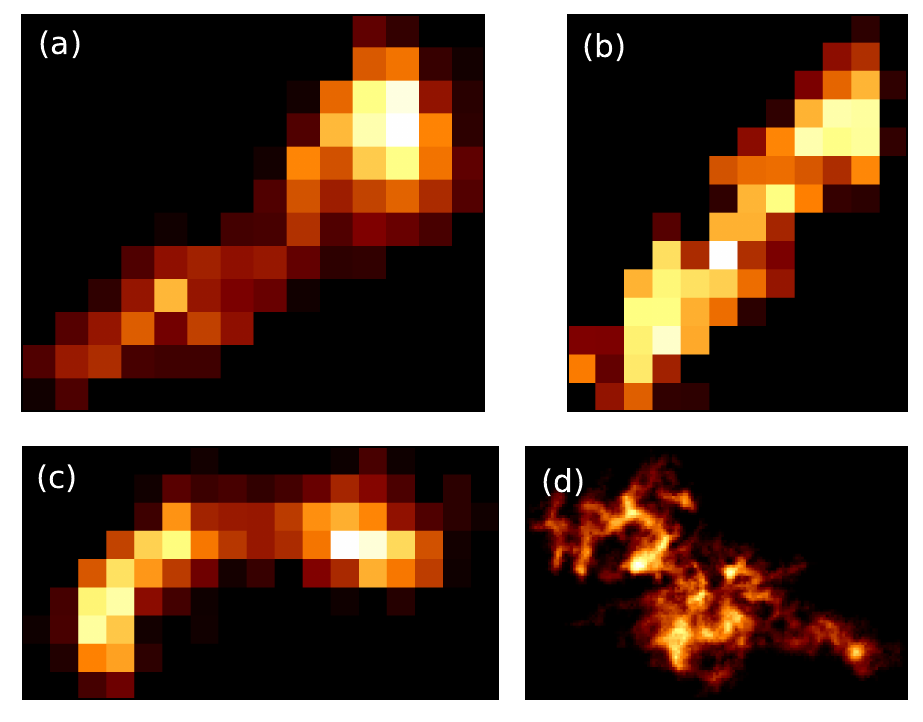}
	\caption{Integrated intensity map of some clouds of the full sample that is included in or excluded from our sample. Panels (a) and (b) show examples of the filamentary molecular clouds that are included in our final sample. Panel (c) shows the curved cloud that we exclude from the final sample, and panel (d) shows the example that is excluded because of large angular scale and network-like substructures.}
	\label{fig1}
\end{figure}

In the following two subsections, we describe the specific steps for selecting the samples through morphology. In Section \ref{Elliptical fitting and aspect ratio selection}, we use elliptical-fitting to exclude those with a small aspect ratio and define a `preliminary' orientation of each cloud with the major axis. In Section \ref{Removal of curved clouds}, we further exclude those that are too curved with respect to this orientation. Fig. \ref{fig1} shows integrated intensity maps of some clouds that are included in or excluded from our sample.

\subsection{Elliptical fitting and aspect ratio selection} \label{Elliptical fitting and aspect ratio selection}
We use elliptical fitting method to retrieve the size and shape of the 18190 molecular clouds of the \citet{2021ApJS..257...51Y} sample. The elliptical fitting results are used for further filtering of the sample.

In this method, the integrated intensity of each pixel within a region is treated as the density of each point on a rigid plate, and the moment of inertia of this plate with respect to its centre of mass can be represented by an elliptical shape.
        
We first calculate the position of the centre of mass:
\begin{equation} 
\begin{split}
    x_{\rm{c}} &=\frac{\int x\rho (x,y)dxdy}{\int \rho (x,y)dxdy},\\
    y_{\rm{c}} &=\frac{\int y\rho (x,y)dxdy}{\int \rho (x,y)dxdy}.
\end{split}
\end{equation}
Then we calculate the moment of inertia tensor:
\begin{equation}
J={\left( \begin{array}{cc}
    J_{\rm{xx}} & J_{\rm{xy}} \\
    J_{\rm{yx}} & J_{\rm{yy}} 
    \end{array}
    \right )},
\end{equation}
with 
\begin{equation}
	J_{\rm{xx}}=\frac{\int (x-x_{\rm{c}})^2\rho (x,y)dxdy}{\int \rho (x,y)dxdy},
\end{equation}
\begin{equation} 
    J_{\rm{yy}}=\frac{\int (y-y_{\rm{c}})^2\rho (x,y)dxdy}{\int \rho (x,y)dxdy},
\end{equation}
\begin{equation}
	J_{\rm{xy}}=J_{\rm{yx}}=\frac{\int (x-x_{\rm{c}})(y-y_{\rm{c}})\rho (x,y)dxdy}{\int \rho (x,y)dxdy}.
\end{equation}
The direction of the two eigenvectors of $J$ corresponds to the direction of the major and minor axes of the two-dimensional intensity distribution under study.
The two eigenvalues describe the length scales in these two directions.

We only include clouds with aspect ratio (the ratio between the lengths of the major and minor axes) greater than 10 in our sample, to ensure that they are visually `slender'. The orientation of a filament is defined as the major axis direction of the elliptical fitting.

\subsection{Removal of curved clouds} \label{Removal of curved clouds}

After the previous two steps of selection based on size and aspect ratio, we get a sample with 817 sources. Some of the selected clouds may still appear curved, for which an orientation is not well-defined. Therefore we take a further step to exclude these curved ones.

We make use of the boundary morphology of a cloud to decide whether it is curved or not (the minimum cut-off is set to $2\sigma$).
We obtain the pixel coordinates of the boundary points from the integrated intensity map of the sample, then we calculate the distance $(d)$ between each boundary point and the major axis of the elliptical fitting. Intuitively, if the sample is more curved, the variance of this distance $d$ will be larger. Therefore, we calculate the variance of the distance over the boundary:
\begin{equation}
	Var(d)=\frac{\sum_{i = 1}^{n} (d_{\rm{i}}-\overline{d})^2}{n},
\end{equation}
where $n$ is the number of boundary points of a cloud.

\begin{figure}
    \centering
    \includegraphics[width=0.9\columnwidth]{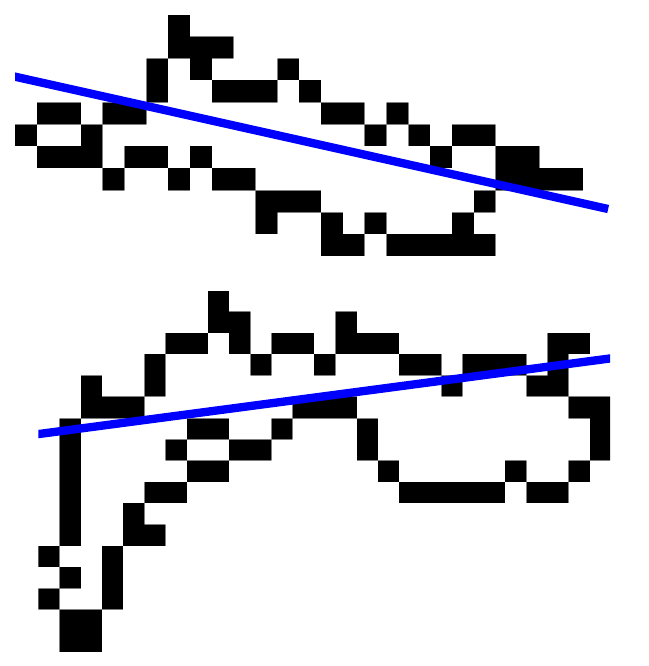}
	\caption{These two cloud silhouettes are examples selected from the filaments obtained by elliptical fitting. The blue line represents the major axes obtained from elliptical fitting, which indicate the orientation of the filament. The top panel shows a straight cloud (included in our final sample), which $Var(d)$=1.069 pixel$^2$, and the bottom panel shows a curved cloud (not included in our final sample), with $Var(d)$=6.021 pixel$^2$. The pixel size here is 30$^{\prime\prime}$.}
	\label{fig2}
\end{figure}

Fig. \ref{fig2} shows examples of straight and curved sample. Calculations of many other filaments also show that $Var(d)$ of curved clouds are greater than those of straight clouds.
To eliminate the influence of the filament's overall size, we define a dimensionless parameter
\begin{equation}
	s=\sqrt{Var(d)/S},
	\label{parameter s}
\end{equation}
where $S$ is the number of pixels within a filament.
We expect that a filament with a small value of $s$ has a straight shape.

\begin{figure*}
    \centering
    \includegraphics[width=2\columnwidth]{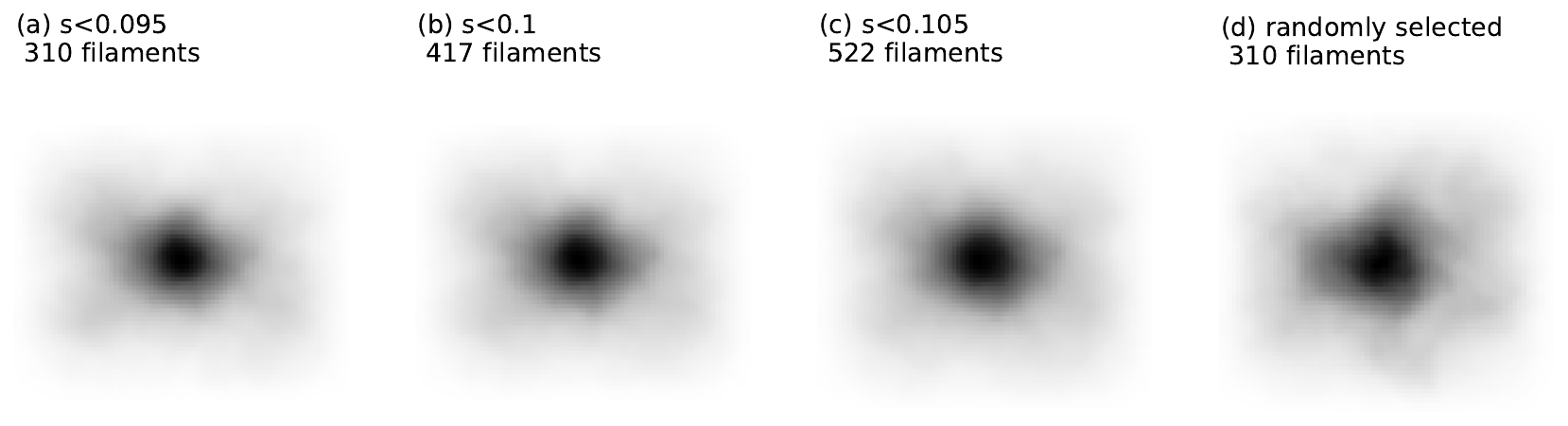}
	\caption{The overlay map of filaments obtained by different thresholds. Panels (a), (b), and (c) show the subsamples obtained using thresholds of 0.095, 0.1, and 0.105, respectively, with the number of samples marked in each panel. Panel (d) presents the overlay map of 310 randomly selected samples.}
	\label{fig3}
\end{figure*}

Since we are unable to define a threshold of parameter $s$ from `first principles' to obtain a suitable sample, we try with three thresholds for it to obtain the three different samples. Then we compare the samples to select one suitable threshold to be used.  To see the effect of using different thresholds, we scale all the filaments in each sample to the same spatial scale and co-add their intensities. The three overlay maps obtained from the three different thresholds are shown in panels (a--c) of Fig. \ref{fig3}.

We also randomly select 310 clouds and find no evidence of a bimodal orientation distribution after stacking them (panel (d) of Fig. \ref{fig3}), thereby confirming the necessity of the step to exclude curved clouds. This test also supports our view that the orientation of curved molecular clouds cannot be accurately described.

With the decrease in the threshold for parameter $s$, the number of sources in the sample also decreases, and we observe that the bimodal orientation distribution of the overlay becomes more pronounced. This confirms our expectation that a lower threshold yields a higher level of straightness and morphological simplicity.

As a compromise between the sample size (for statistical significance) and sample `sharpness' (for morphological straightness), for subsequent studies, we use the sample comprising 522 filaments that correspond to $s<0.105$.

In Appendix \ref{Test of the method of removal of curved clouds}, we also conduct an experiment to confirm that this step does not introduce bias to the sample selection.

\subsection{Estimation of physical parameters of filament}\label{Estimation of physical parameters of filament}

We employ the Parallax-Based Distance Calculator V2 \citep{2016ApJ...823...77R} to estimate the distances of filaments. This tool uses a source's Galactic coordinates and $V\rm _{LSR}$ to assign it to a portion of a spiral arm based on its proximity to spiral arms seen in CO or H\,{\textsc i} surveys. As the distances to a large portion of the spiral arms are now known, we can obtain reliable distance estimations through this model.

\begin{figure}
    \centering
    \includegraphics[width=\columnwidth]{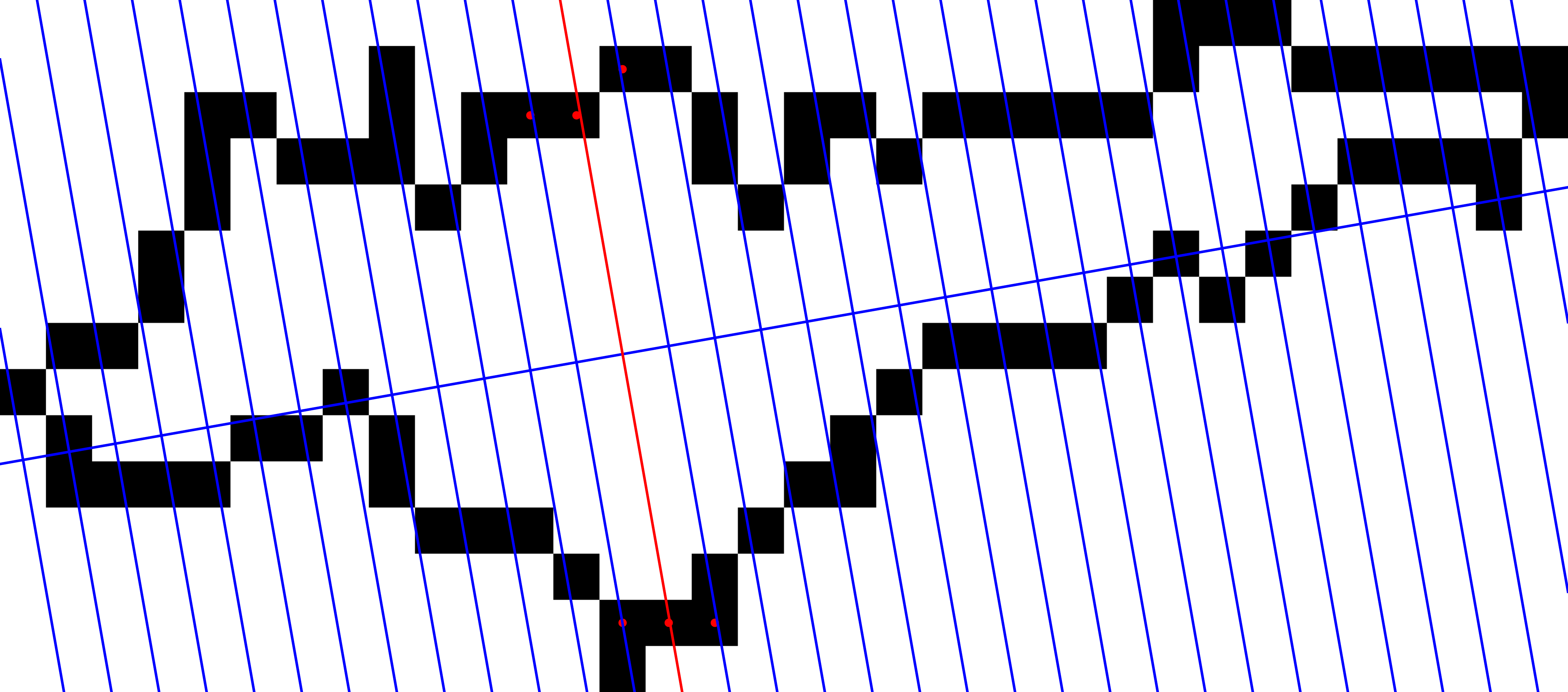}
    \caption{Schematic diagram illustrating the calculation of the average width. The red line is used as an example to illustrate the calculation process, its direction is perpendicular to the orientation of the filament. The pixel size here is 30$^{\prime\prime}$.}
    \label{fig4}
\end{figure}

We further calculate the physical lengths and widths of the filaments using the distances, angular lengths and angular widths.
As shown in Fig. \ref{fig4}, we create a series of lines along the major axis, perpendicular to the orientation of the filament, and employ each line for width calculation. For example, we take the red line in Fig. \ref{fig4} and identify the two closest points to the line. Subsequently, we compute the separation between these two points for each line parallel to the red line. Any separation that is less than or equal to 2 pixels is excluded. Finally, we calculate the average value of all the remaining separations to represent the average width of the filament. We apply the same method to calculate the average length of the filaments.

The velocity width of each filament is obtained through Gaussian fitting to the integrated line profile. The full-width at half maximum ($\text{FWHM} = 2\sqrt{\ln2}\sigma$) of the fitting results are later used to investigate the possible correlation between velocity width and orientation in Section \ref{Correlation of the orientation with the other parameters}.

In Table \ref{Table B1} of Appendix \ref{Parameters of filaments we used}, we provide a list of relevant parameters, including the source name, orientation, coordinates, velocity, estimated distance, physical scale, velocity width, mass, maximum column density, and average column density for the 522 filaments in our sample.

\section{Bimodal orientation distribution of the sample} \label{Bimodal orientation distribution of the sample}
We use the major axis direction of elliptical fitting to define the orientation of filaments in our final sample and study the distribution of the orientation. To estimate the error in the obtained orientation angle, we add Gaussian noise with a standard deviation of $\sqrt{n}\sigma\Delta v$ (where the `$\sigma $' is the RMS noise of the MWISP data, `$\Delta v$' is velocity resolution of the data, and `$n$' is the number of velocity channels (for calculating integrated intensity) of the filaments) to each pixel of the filament, then use elliptical fitting to calculate the orientation. This process is repeated 1000 times to obtain the standard deviation of these orientation data. This procedure is performed for each filament, from which the standard deviation of the orientation angle of each filament is calculated. The errors so calculated for each filament are different, and typically lies in the range $0.2^\circ \to 7.1^\circ$, and we report $1.6^\circ$ as the most probable value. Additionally, we investigate the correlation between the orientation and other physical parameters of the filaments.

\subsection{Bimodal orientation distribution}
\begin{figure}
    \centering
    \includegraphics[width=\columnwidth]{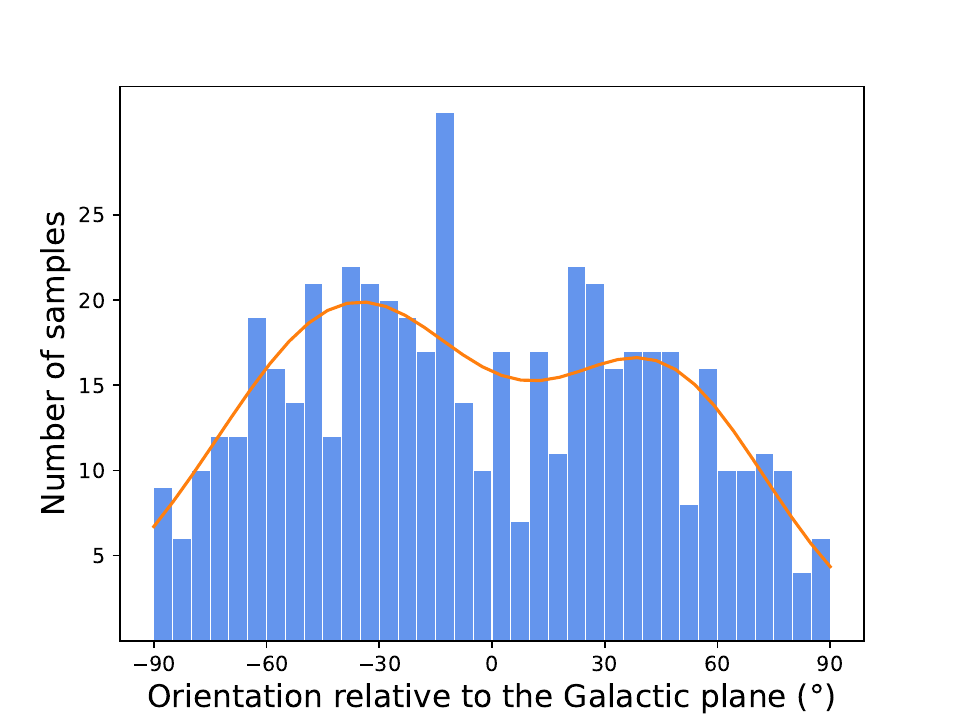}
    \caption{The distribution of the orientation angles. The curve is the fitted double Gaussian function.}
    \label{fig5}
\end{figure}

Fig. \ref{fig5} shows the distribution of the angle between the orientation of filaments in our final sample and the Galactic plane. We can see that the orientation of the filaments mainly concentrates on two ranges. To obtain the parameters of the bimodal orientation distribution, we use a double Gaussian function to fit the histogram. The centres of the two Gaussian functions are $-38.1^\circ $ and $42.0^\circ $, and the 1$\sigma$ of the two Gaussian functions are $35.4^\circ $ and $27.4^\circ $.

\begin{figure}
    \centering
    \includegraphics[width=1\columnwidth]{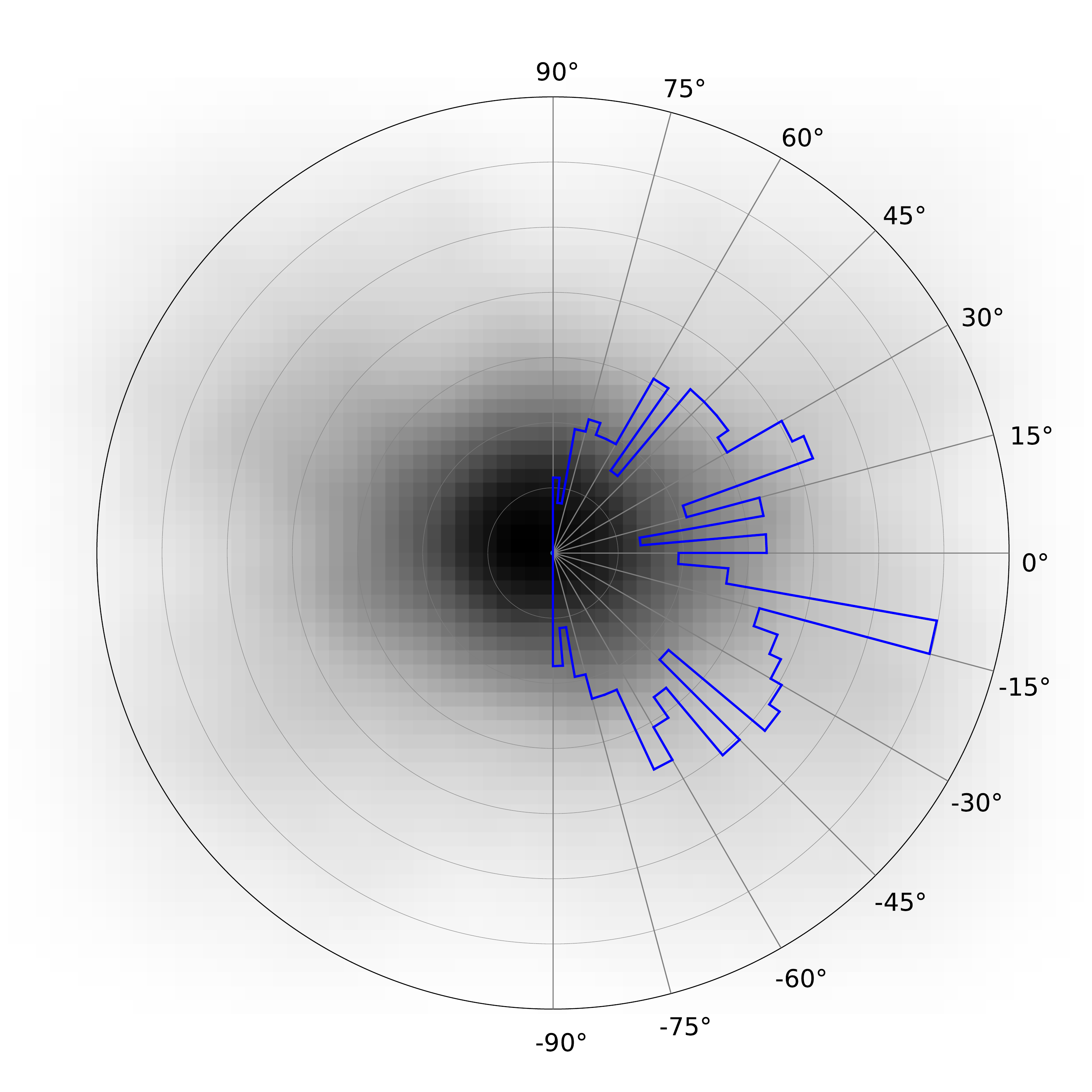}
    \caption{Polar histogram and overlay map of the filaments. The grayscale is created by adding the intensity of all the clouds (scaled to the same length) together. This figure is an alternative to Fig. \ref{fig5} for showing the orientation distribution.}
    \label{fig6}
\end{figure}

As an intuitive way to show the orientation distribution and an alternative to Fig. \ref{fig5}, we make an overlay map in Fig. \ref{fig6}, where the background, like in Fig. \ref{fig3}, is obtained using the method described in Section \ref{Removal of curved clouds}. The histogram shows the orientation distribution in polar coordinates.
Both the overlay map and the polar histogram reveal a bimodal orientation distribution.

To see whether the relative orientation between two filaments depends on the their spatial separations, we define the following function
\begin{equation}
f(\ell,\theta)d\ell d\theta \equiv \frac{\#\{(i,j)|\ell\le d_{ij}\le\ell+d\ell\,\land\, \theta\le\angle_{ij}\le\theta+d\theta\}}{\#\{(i,j)|\ell\le d_{ij}\le\ell+d\ell\}},
\end{equation}
where the symbol $\#$ means to take the number of elements of a set, $\land$ means `logical and', $d_{ij}$ is the (angular) distance between filament $i$ and $j$, and $\angle_{ij}$ is the relative angle between the two.
Thus $f(\ell,\theta)d\ell d\theta$ quantifies the fraction of filament pairs with angular separations between $\ell$ and $\ell+d\ell$ having relative orientation between $\theta$ and $\theta+d\theta$.

We calculate $f(\ell,\theta)d\ell d\theta$ as follows.
We first calculate the relative orientations between each pair of filaments and obtain a total of 135981 ($=522{\times} 521/2$, where $522$ is the number of filaments in our sample) relative orientations. We then divide the pairs into intervals of $5^\circ$ based on their angular separations, and divide the relative orientations ($0^\circ$ to $90^\circ$) into six intervals. Finally, we calculate the fraction of the relative orientations in each separation interval.

\begin{figure}
    \centering
    \includegraphics[width=1\columnwidth]{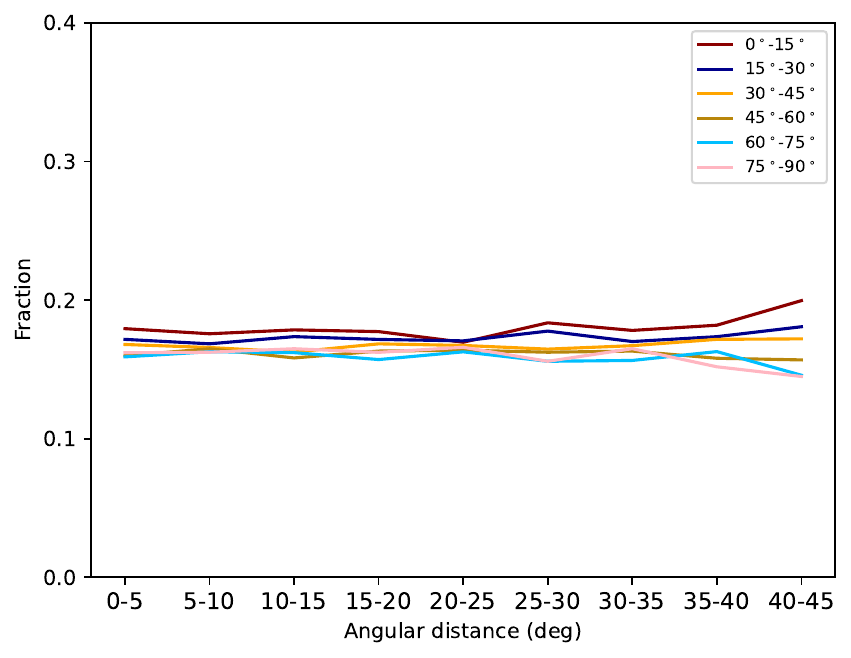}
    \caption{The fraction of different relative orientations for each angular separation interval. The six lines represent the relative orientations corresponding to the six intervals indicated by the labels, and the horizontal axis represents the corresponding angular separation.}
    \label{fig7}
\end{figure}

Fig. \ref{fig7} shows the fraction of different relative orientations for each angular separation interval. We observe no correlation between the relative orientation and the angular separation of the filaments.

\begin{figure*}
    \centering
    \includegraphics[width=2\columnwidth]{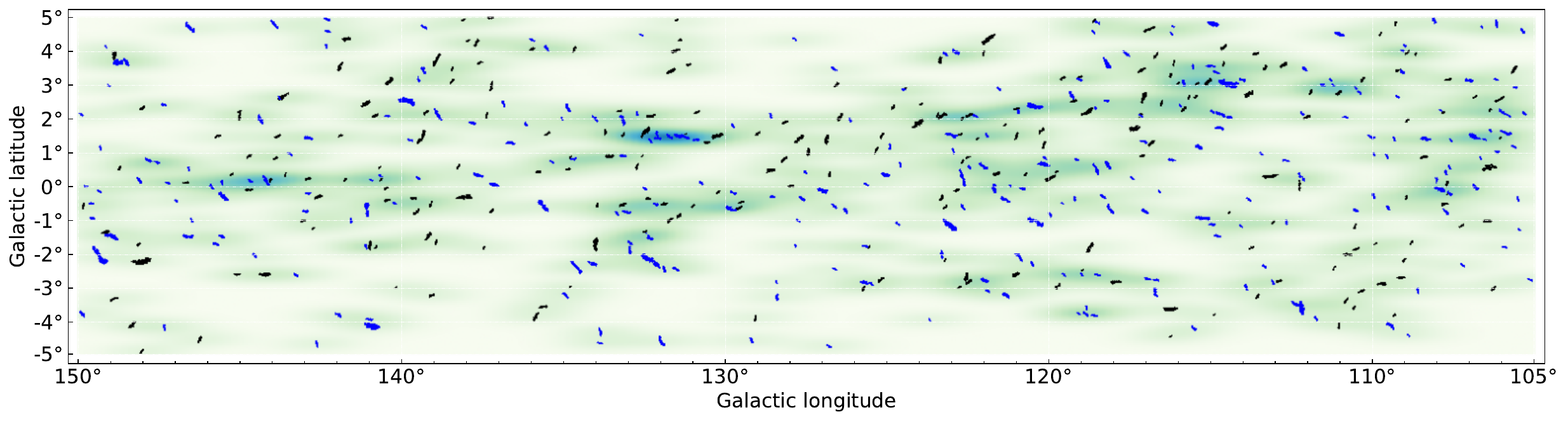}
    \caption{Distribution of all filaments in our final sample in Galactic coordinates. Each filament is coloured in black (for orientation angle $>0$) or blue (for orientation angle $<0$). The background colour indicates the number density of filaments in our sample. The bimodal orientation distribution is already visually perceivable in this figure.}
    \label{fig8}
\end{figure*}

The distribution of all the filaments in our final sample in plane of the sky is shown in Fig. \ref{fig8}.
The background colour scale is a density map of filament count, generated using the scipy function \texttt{gaussian\_kde}\footnote{\url{https://docs.scipy.org/doc/scipy/reference/generated/scipy.stats.gaussian_kde.html}}, where darker colours indicate higher density.
The bimodal orientation distribution is already noticeable `by eye' from this figure, and apparently the orientation does not appear to be correlated with spatial location.

\subsection{Correlation of the orientation with the other parameters}\label{Correlation of the orientation with the other parameters}

As an attempt to understand the physical origin of the bimodal distribution, we look for any possible correlation between the orientation and other physical parameters of the filaments, such as Galactic longitude and latitude, radial velocity, physical scale, and velocity width (see Section \ref{Estimation of physical parameters of filament}).

\begin{figure*}
    \centering
    \includegraphics[width=2\columnwidth]{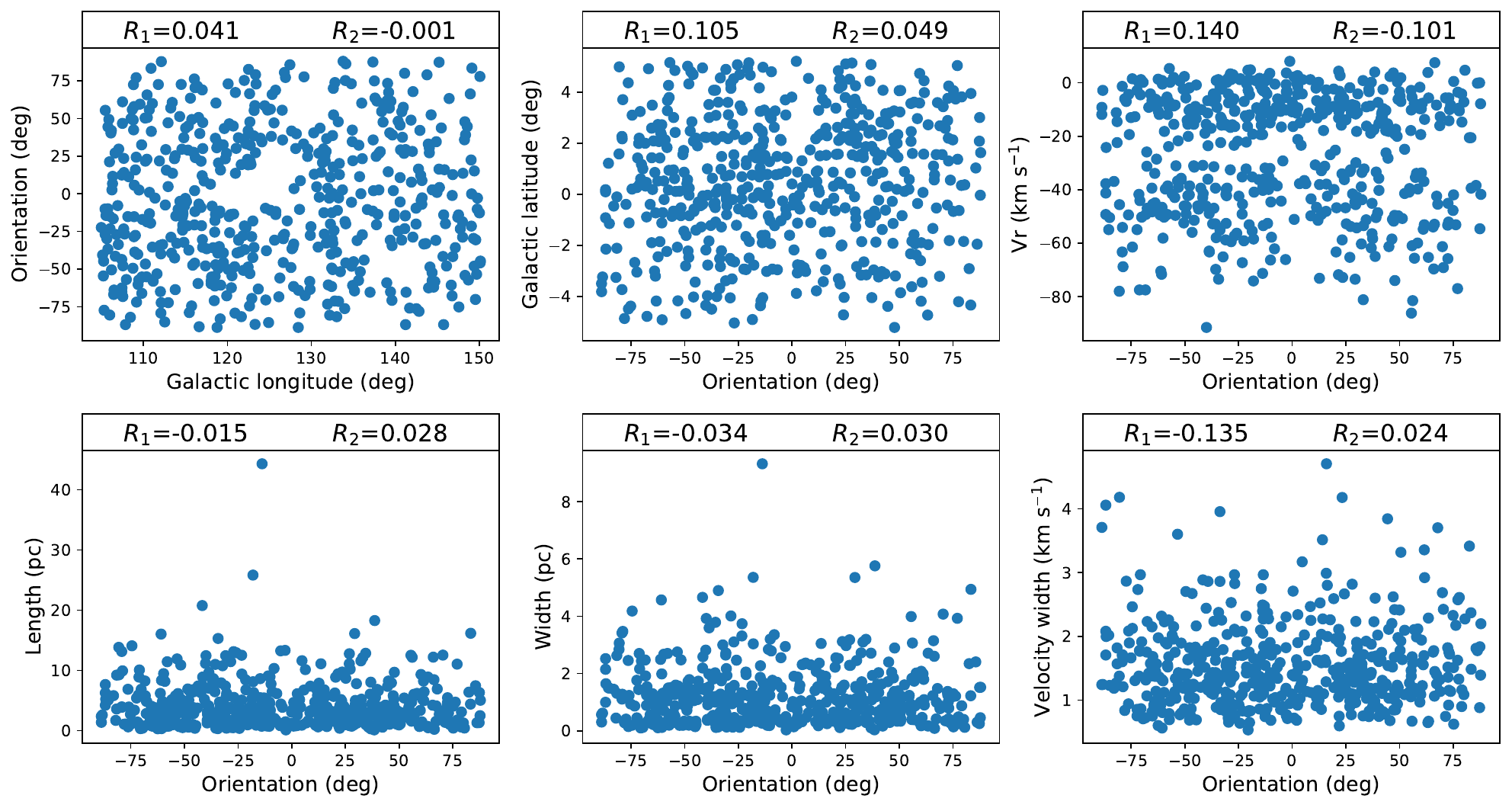}
    \caption{The (lack of) correlation between orientation and other physical parameters of the filaments in our final sample. $R_1$ is the correlation coefficient obtained from sources with nagative orientation angles, and $R_2$ from sources with positive orientation angles.}
    \label{fig9}
\end{figure*}

The scatter plots of these parameters \textsl{versus} orientation angle are shown in Fig. \ref{fig9}, where in each panel the parameters $R_1$ and $R_2$ are the correlation coefficients for sources with negative or positive orientation angle, respectively.
The maximum absolute value of these correlation coefficients never exceeds 0.15, thus we do not find any obvious correlation between orientation and other physical parameters, and the origin of the bimodal orientation distribution remains a puzzle at this moment.
    
\section{Head-tail asymmetry} \label{Head-tail asymmetry}
For the filaments in our sample, we observe that some have a morphological asymmetry, that is, one end is `thicker' than the other (See, e.g., panel (a) in Fig. \ref{fig1}). One possible explanation for the observed accumulation of matter at the end of filaments is the `edge effect' \citep{1983A&A...119..109B}, caused by the filament's self-gravity.
This effect has been predicted by theory, and has been observationally found \citep{2013A&A...554L...2Z,2020ApJ...899..167B,2020A&A...637A..67Y}.
In this work, we use a large sample from the MWISP survey to statistically analyse the head-tail asymmetry of filamentary molecular clouds.

\subsection{One-dimensional intensity distribution along the major axis and the number of peaks} \label{One-dimensional integrated intensity distribution along the major axis}

\begin{figure}
    \centering
    \includegraphics[width=0.9\columnwidth]{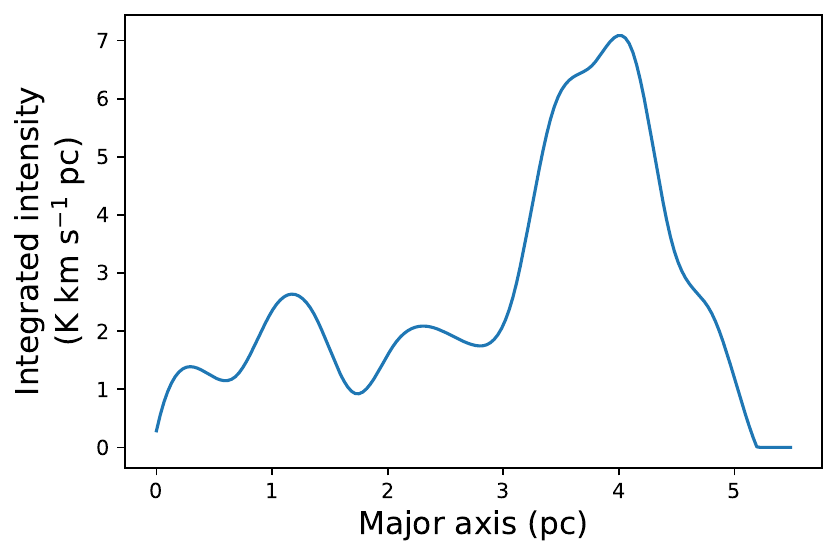}
    \caption{An example of one-dimensional integrated intensity profile along the major axis.}
    \label{fig10}
\end{figure}

We quantify this head-tail asymmetry using the integrated intensity along the major axis of the filaments.
Fig. \ref{fig10} shows one example for the one-dimensional integrated intensity profile.

The number of peaks in the curve represents the concentration of integrated intensity, which can influence the asymmetry of the curve. Therefore, we perform statistical analysis to study the head-tail asymmetry in filaments with an equal number of peaks.

To determine the number of peaks in the curve, we perform multi-Gaussian fitting on the integrated profile and extract a series of coordinates for candidate peaks from the centres of the multi-Gaussian functions. The peak number of the curve is defined based on the parameters of the Gaussian function. We exclude candidate peaks that have peak values less than $3\sqrt{n}\sigma $, where the `$\sigma $' is the RMS noise of the MWISP data, and `$n$' is the number of integration steps, which includes two components: the number of pixels in the velocity dimension and the number of pixels in the direction perpendicular to the filament's major axis before the interpolation. Next, we calculate the separation between the two closest candidate peaks along the major axis. We use $0.85\text{FWHM}$ of the Gaussian fitting to a peak to represent its width and compare it with the separation of the two peaks. If both widths of the two peaks are less than the separation, we consider that these two candidate peaks are two independent peaks.

\begin{figure}
    \centering
    \includegraphics[width=\columnwidth]{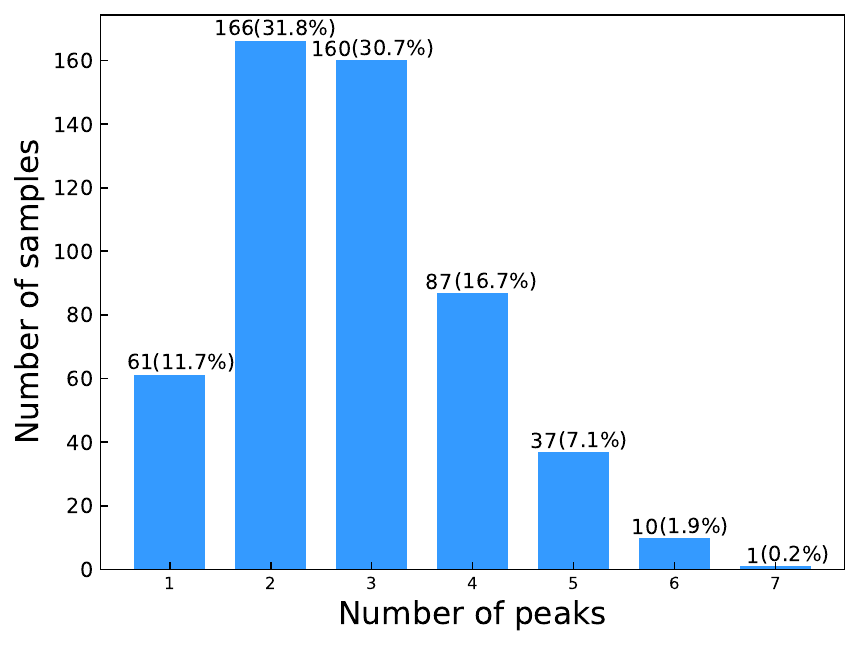}
    \caption{Distribution of the number of peaks of the filaments in our final sample.}
    \label{fig11}
\end{figure}

In our sample, 11.7 per cent of the filaments' one-dimensional integrated intensity profiles are single-peaked, 31.8 per cent are double-peaked, 30.7 per cent have three peaks, and the fraction of filaments with four or more peaks is $\approx$26 per cent (see Fig. \ref{fig11}). 

Subsequently, we analyse the filaments based on different peak numbers. Since the limited number of samples with five or six peaks, we only construct subsamples for filaments containing four peaks or fewer.

\subsection{Head-tail asymmetry of the integrated intensity}\label{Head-tail asymmetry of the integrated intensity}

We use two approaches to quantify the asymmetry of the integrated intensity curves of the filaments.
We first use skewness to quantify the degree of asymmetry in the one-dimensional integrated intensity. Skewness is a numerical measure of the degree of asymmetry of a probability density distribution curve, with positive and negative skewness representing left and right deviation of the data, defined as the third-order standardized moment of the random variable:
\begin{equation} \label{skewness}
    Skew(X) =\frac{\frac{1}{n}\sum_{i = 1}^{n}(x_{\rm{i}}-\bar x)^3}{[\frac{1}{n-1}\sum_{i = 1}^{n}(x_{\rm{i}}-\bar x)^2]^{3/2}}.
\end{equation}
If we consider the value of each point on the integrated intensity profile as a probability density, the skewness can be used to describe the head-tail asymmetry of the curve.

Alternatively, we define a parameter $p$ to quantify the degree of asymmetry in a curve.
We calculate two moments of inertia of the cloud relative to two axes parallel to the minor axis and located at the two endpoints of the major axis:
\begin{equation}
\begin{split}
    I_1 &=\int_{0}^{L} r^2\rho (r) \,dr,\\
    I_2 &=\int_{0}^{L} (L-r)^2\rho (r) \,dr,
\end{split}
\end{equation}
and the parameter $p$ is defined as:
\begin{equation}
    p=\ln \left[\frac{\max (I_1,I_2)}{\min (I_1,I_2)} \right].
\end{equation}
Thus a higher value for $p$ means the one dimensional distribution along the major axis is more asymmetric.
To demonstrate the morphological difference of filaments with different $p$, we present $^{12}$CO $(J=1-0)$ velocity-integrated intensity maps of representative filaments in increasing order of $p$ in Appendix \ref{example images of filaments}.  A visual inspection there indicates that filaments with $p\gtrsim0.35$ may be reasonably considered to be `obviously head-tail asymmetric'.

\begin{figure*}
\centering
\includegraphics[width=2\columnwidth]{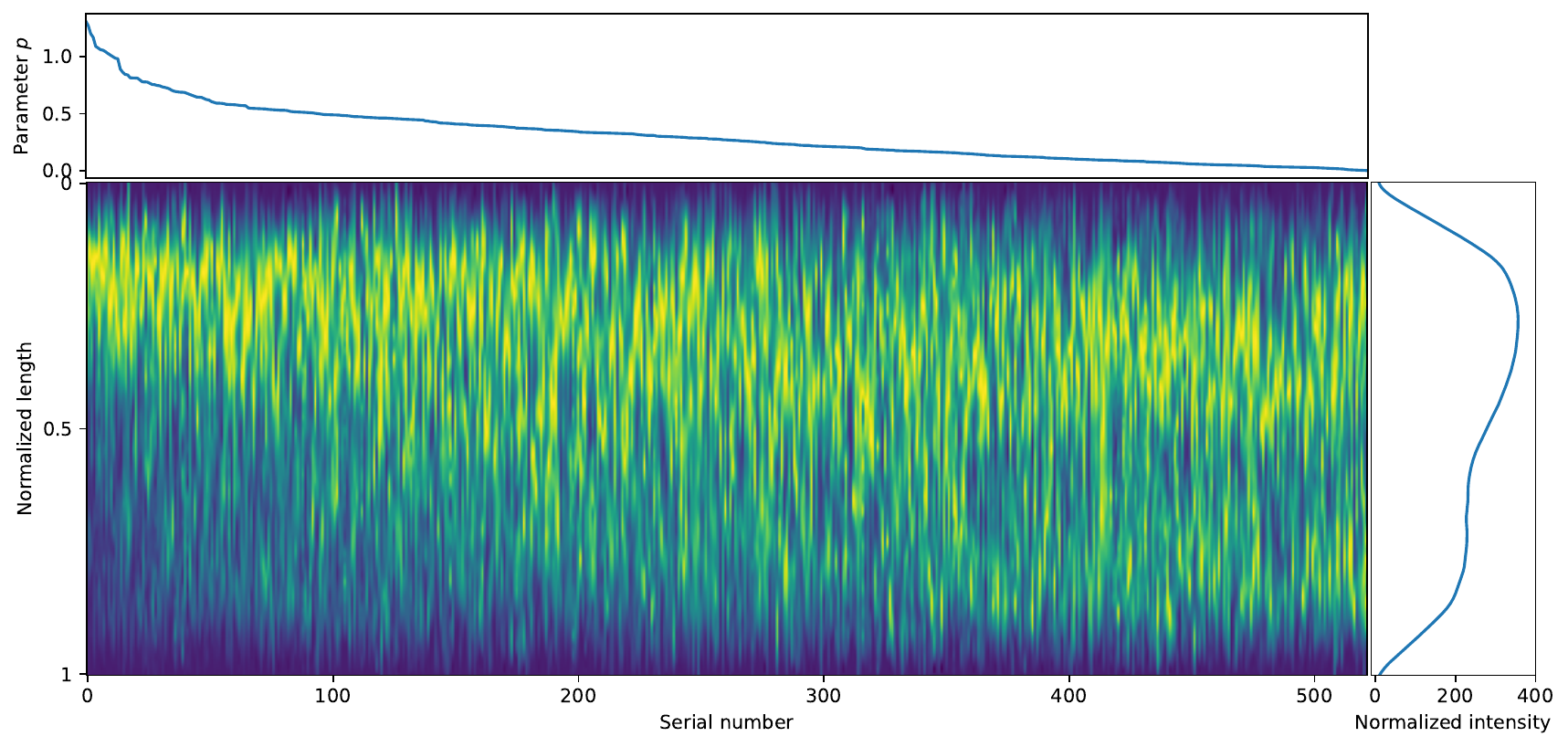}
\caption{The colour shows the intensity distribution along the major axis ($y$ axis, with length normalized to 1) of the filaments in our sample.  The filaments are sorted by the parameter $p$ (top panel), and are oriented such that the `heavier' half are on the upper side.  The right panel shows a sum of the intensity profile of all the filaments.}
\label{fig12}
\end{figure*}

After calculating the parameter $p$ for all the filaments in the sample, we sort them in descending order of $p$. We then normalize the length and one-dimensional integrated intensity of all filaments and adjust their orientation so that the `heavier' end is aligned. Using the serial number of filaments as the x-axis and the normalized length as the y-axis, we draw the intensity distribution and co-add the normalized intensities of all filaments to obtain the one-dimensional superimposed integrated intensity curve (Fig. \ref{fig12}).
It can be seen that as the parameter $p$ decreases, the intensity of the filaments gradually transitions from asymmetrical to symmetrical, and the overall intensity distribution along the major axis indeed tends to be one-sided.

\begin{figure*}
        \centering
        \includegraphics[width=2\columnwidth]{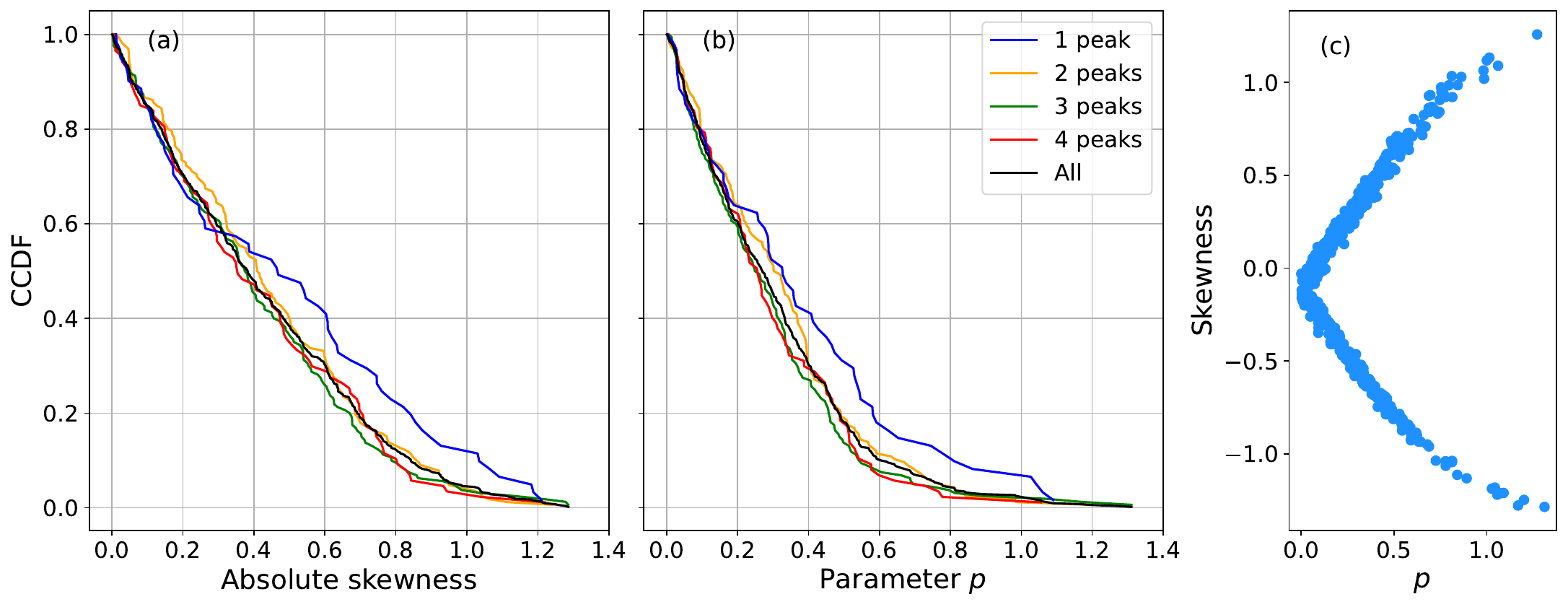}
        \caption{Complementary cumulative distribution function (CCDF) of the absolute skewness and the parameter $p$. Panel (a) shows the CCDF of absolute skewness, while panel (b) shows the CCDF of parameter $p$, for filaments with 1--4 peaks and the whole sample.  Panel (c) illustrates the correlation between the skewness and the parameter $p$.}
        \label{fig13}
    \end{figure*}

The complementary cumulative distribution function (CCDF) of skewness and $p$ are shown in panels (a) and (b) of Fig. \ref{fig13}. Panel (c) of that figure shows the correlation between skewness and $p$, which demonstrates that both can be used to characterize the asymmetric distribution.
The CCDF plot shows that about 40 per cent of all the filaments in our sample have $p>0.35$, so we may say that about 40 per cent of our sample are `obviously' head-tail asymmetric.
Filaments with a single peak tend to be more asymmetric (having a higher value of skewness or $p$) than those with more peaks, which is not unexpected.\footnote{In a grossly simplified picture in which the peaks are thought of as point mass randomly distributed within an otherwise massless filament, in the case of single peak the expected value of $p$ is $\ln16\simeq2.77$, and in the case of two peaks the expected value of $p$ is $\simeq0.53$.}

\subsection{Correlation of the parameter $p$ with other physical parameters}\label{Correlation of the parameter p with the other parameters}

To understand whether the degree of head-tail asymmetry is related to the evolution of filaments, we analyse the relationship between parameter $p$ and other parameters of filaments, including physical scale, velocity width, mass, maximum column density, and orientation.

The column densities and masses are simply calculated using the $X$-factor approach, with $X_{\text{CO}}=2 {\times} 10^{20} \,\text{cm}^{-2}(\text{K km s}^{-1})^{-1}$ and a $\pm$30 per cent uncertainty, as recommended by \cite{2013ARA&A..51..207B}:
\begin{equation}
\begin{split}
    \text{N}(\text{H}_2)&=X_{\text{CO}}\text{W}(\text{CO}), \\
    \text{Mass}&=\sum \text{N}(\text{H}_2)\times \Delta\Omega\times d^2,
\end{split}
\end{equation}
where the sum is over all the pixels of a filament, $\text{W}(\text{CO})$ is the integrated line intensity of $^{12}\text{C}^{16}\text{O}\ ({J=1-0})$, $\Delta\Omega$ is the angular area of a pixel, and $d$ is the distance to the filament.

\begin{figure*}
    \centering
    \includegraphics[width=2\columnwidth]{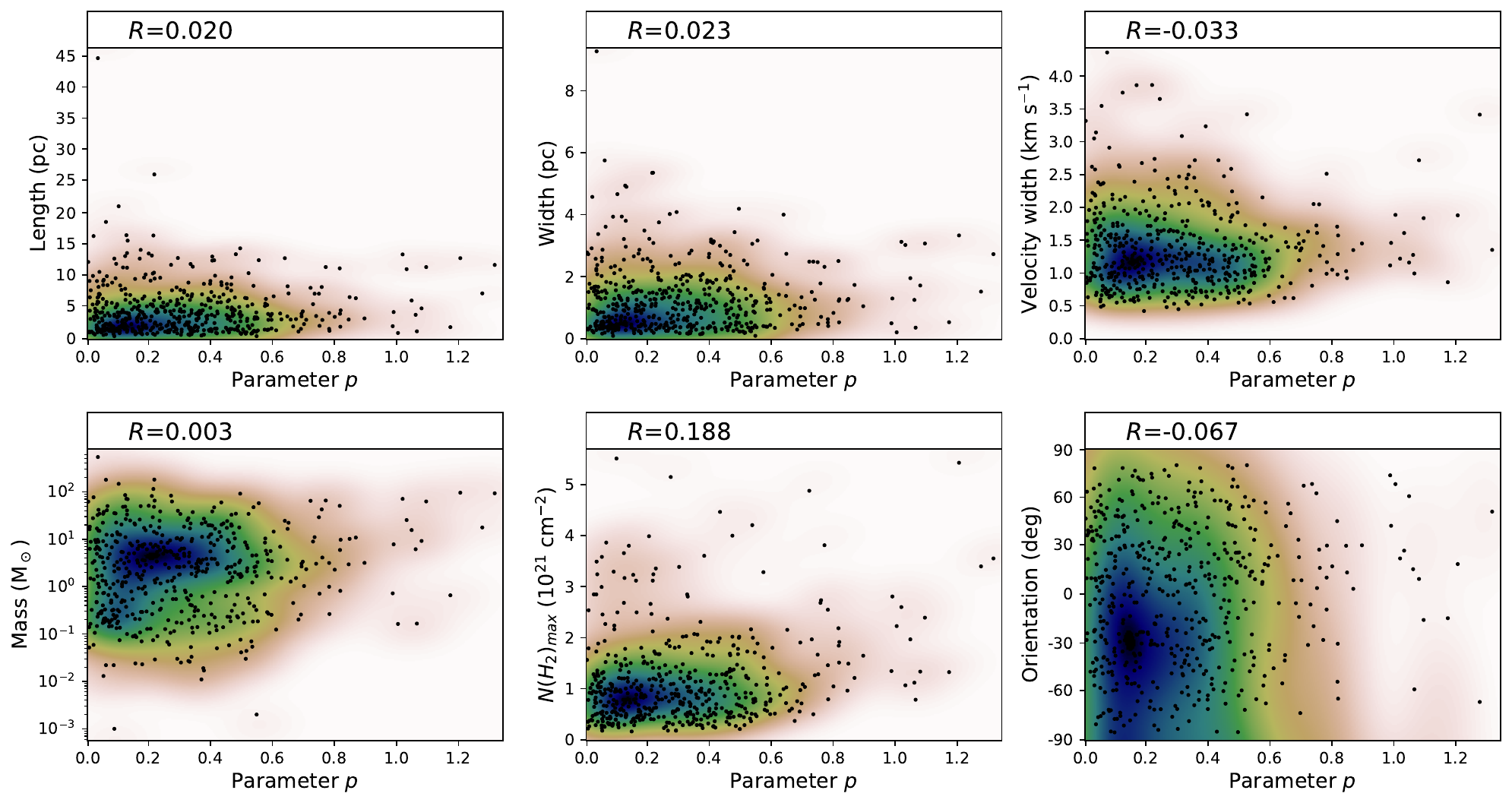}
    \caption{The (lack of) correlation between parameter $p$ and other physical parameters of the filaments in our final sample. The background colour indicates the number density of the scatter points.}
    \label{fig14}
\end{figure*}

We do not see any significant correlation between parameter $p$ and other physical parameters in the resulting scatter plot Fig. \ref{fig14}, while we do notice that the maximum column density, $N(\text{H}_2)_\text{max}$, seems to be weakly correlated with $p$.
This could be understood if the very head-tail asymmetric filaments have been undergoing mass concentration towards one end, leading to the elevated column density.

\section{Conclusions and Discussions} \label{Conclusion}
In this work, we use part of the $^{12}$CO $(J=1-0)$ data from the MWISP project to study the orientation and head-tail asymmetry of filaments with a simple straight morphology.
We construct a series of samples during the process of excluding curved clouds. From the results, a distinct bimodal orientation distribution is seen (Figs. \ref{fig5} and \ref{fig6}), and the bimodal orientation becomes more pronounced as the screening criteria become more stringent (while still ensuring the inclusion of a certain amount of clouds).
We also find that certain filamentary molecular clouds exhibit distinct head-tail asymmetry, with one end being wider or brighter than the other.
The main conclusions are as follows:
\begin{enumerate}
\item We find a bimodal orientation distribution of the sample. By employing a double Gaussian fitting, we obtain that the centres of the two Gaussian functions are $-38.1^\circ$ and $42.0^\circ$, and the 1$\sigma$ of the two Gaussian functions are $35.4^\circ$ and $27.4^\circ$, respectively.

\item Most filaments exhibit a small number of peaks. Specifically, 43.5 per cent of the filaments' one-dimensional integrated intensity profiles show either a single peak or double peaks, while $\sim26$ per cent of filaments have four or more peaks.  We note that the number of peaks could be affected by the spatial resolution of the data.

\item Based on the complementary cumulative distribution function (CCDF) of the two asymmetric parameters, absolute skewness and $p$, we find that filaments with a single peak tend to be more asymmetrical than those with more peaks, and about 40 per cent of the whole sample have obvious head-tail asymmetry.
\end{enumerate}

The orientation of filaments can be influenced by various factors, including the local environment such as the magnetic field, physical processes within the cloud, and their formation and evolution. Since turbulence can compress gas in any direction, we may use the bimodal orientation distribution of filaments to exclude the possibility that turbulence solely dominates the formation of filaments.

The data used in this work are in the form of a data cube in position-position-velocity (ppv) space, rather than position-position-position (ppp) space. The impact of projection effects cannot be eliminated. If the three-dimensional orientations of filaments in the sample could be obtained, their distribution might indeed be entirely different. The filaments might not even exhibit a filamentary structure. However, this does not negate the fact that the two-dimensional projection of orientations exhibits bimodality.

Several studies on filaments have concluded that filaments are oriented along the Galactic plane. However, it is noteworthy that these studies focus on giant molecular filaments (GMFs). \cite{2014ApJ...797...53G} find that the very long and thin infrared dark cloud `Nessie' \citep{2010ApJ...719L.185J} lies not just parallel to the Galactic plane, but also \emph{in} the Galactic plane. \citet{2015ApJ...815...23Z} discovered additional `bones' resembling `Nessie', aligned parallel to the Galactic plane. However, \cite{2021A&A...651L...4S} found that most of the filamentary structures in the $^{12}$CO and $^{13}$CO emission do \emph{not} show a global preferential orientation either parallel or perpendicular to the Galactic plane, and they propose that a selection bias may be introduced by the GMF identification in the general trends found for the GMFs, which appear to be aligned with the Galactic plane.

Our work differs from these previous studies in a few aspects. Firstly, in the sample selection, we focus on small-scale filaments with simple straight morphology, as opposed to large-scale filaments (GMFs). Secondly, in the orientation calculation, we derive orientations from the major axis direction through elliptical fitting, while previous studies typically employed methods based on ideas from computer graphics. For example, \cite{2013ApJ...774..128S} and \cite{2021A&A...651L...4S} employ methods based on the direction perpendicular to the local intensity gradient and the direction characterized by the eigenvector of the Hessian matrix, respectively, to statistically analyze the orientation of the density structures.  In Appendix \ref{Cross-check with different analysis method}, we also applied the method of \cite{2013ApJ...774..128S} to our data to cross-check our results.  Finally, the distribution of filament orientation is expected to vary across different regions of the Galaxy, and our work only focuses on the molecular clouds in the second Galactic quadrant. All these factors contribute to the disparities between our and previous results.

\cite{2013MNRAS.436.3707L} found that filaments in the Gould Belt tend to align either parallel or perpendicular to the local mean magnetic field directions. They interpreted this as evidence that the magnetic fields within the intercloud medium are strong enough to influence gravitational contraction, resulting in the formation of flattened condensations oriented perpendicular to the magnetic fields, and channeling the turbulence and leading to the alignment of filaments along the magnetic fields.
\citet{2016A&A...586A.138P} performed a statistical analysis to assess the relationship between the magnetic field projected onto the plane of the sky ($\bm{B}_\bot$) and the gas column density ($N_\text{H}$) structures. They observed a systematic change in the relative orientation between $\bm{B}_\bot$ and $N_\text{H}$ structure across different regions, with the orientation being parallel in areas with the lowest column density and perpendicular in areas with the highest column density.

While in the current paper we did not notice any correlation between orientation and other physical parameters, we do notice hints of presence of a global pattern of the filament orientations (Fig. \ref{fig8}).
In Appendix \ref{secAppendixSinusoidal} we show that a global sinusoidal arrangement of the filaments could lead to a bimodal orientation distribution.
In our next study, we will investigate the potential correlation between the local magnetic field and the orientation of the filaments.

The head-tail asymmetry of filaments may be explained by the `edge effect' that has been previously studied. \cite{2020A&A...637A..67Y} conducted an investigation on filament S242, which contains concentrated massive clumps and YSO clusters at its end regions. They proposed a two-stage fragmentation scenario for filament S242. Firstly, gravitational focusing leads to the preferential concentration of material at the end regions, resulting in a significant acquisition of mass by the end-clumps. Secondly, longitudinal accretion becomes dominant in the vicinity of the high-mass clumps.

This `edge effect' explains the accumulation of material at both ends of a filament and has been observed in several studies. However, our current study emphasizes the concentration of matter towards one end rather than both ends as described by the `edge effect'.
The head-tail asymmetry of filaments may indicate different physical conditions at the two ends of a filament.
By further investigating this asymmetry and its correlation with the local environment, we can deepen our understanding of the structure and evolution of molecular clouds.

\section*{Acknowledgements}
This work makes use of the data from the Milky Way Imaging Scroll Painting (MWISP) project. MWISP was sponsored by the National Key R\&D Program of China with grant 2017YFA0402701, and the CAS Key Research Program of Frontier Sciences with grant QYZDJ-SSW-SLH047. We are also grateful to the Parallax-Based Distance Calculator V2 tool for its help in estimating the distances of the filaments. F.Du and W.Ge are supported by NSFC through grants 12041305 and 11873094, and National Key R\&D Program of China through grant 2023YFA1608000. This work is also supported by China Manned Space Program through its Space Application System.

\section*{Data Availability}
The data used in this work have been published on the Science Data Bank. \href{https://www.scidb.cn/en/s/jmmeme}{https://www.scidb.cn/en/s/jmmeme}


\newpage
\bibliographystyle{mnras}

\begin{thebibliography}{}
    \makeatletter
    \relax
    \def\mn@urlcharsother{\let\do\@makeother \do\$\do\&\do\#\do\^\do\_\do\%\do\~}
    \def\mn@doi{\begingroup\mn@urlcharsother \@ifnextchar [ {\mn@doi@}
      {\mn@doi@[]}}
    \def\mn@doi@[#1]#2{\def\@tempa{#1}\ifx\@tempa\@empty \href
      {http://dx.doi.org/#2} {doi:#2}\else \href {http://dx.doi.org/#2} {#1}\fi
      \endgroup}
    \def\mn@eprint#1#2{\mn@eprint@#1:#2::\@nil}
    \def\mn@eprint@arXiv#1{\href {http://arxiv.org/abs/#1} {{\tt arXiv:#1}}}
    \def\mn@eprint@dblp#1{\href {http://dblp.uni-trier.de/rec/bibtex/#1.xml}
      {dblp:#1}}
    \def\mn@eprint@#1:#2:#3:#4\@nil{\def\@tempa {#1}\def\@tempb {#2}\def\@tempc
      {#3}\ifx \@tempc \@empty \let \@tempc \@tempb \let \@tempb \@tempa \fi \ifx
      \@tempb \@empty \def\@tempb {arXiv}\fi \@ifundefined
      {mn@eprint@\@tempb}{\@tempb:\@tempc}{\expandafter \expandafter \csname
      mn@eprint@\@tempb\endcsname \expandafter{\@tempc}}}
    
    \bibitem[\protect\citeauthoryear{{Abergel}, {Boulanger}, {Mizuno}  \&
      {Fukui}}{{Abergel} et~al.}{1994}]{1994ApJ...423L..59A}
    {Abergel} A.,  {Boulanger} F.,  {Mizuno} A.,   {Fukui} Y.,  1994, \mn@doi
      [\apjl] {10.1086/187235}, \href
      {https://ui.adsabs.harvard.edu/abs/1994ApJ...423L..59A} {423, L59}
    
    \bibitem[\protect\citeauthoryear{{Andr{\'e}} et~al.,}{{Andr{\'e}}
      et~al.}{2010}]{2010A&A...518L.102A}
    {Andr{\'e}} P.,  et~al., 2010, \mn@doi [\aap] {10.1051/0004-6361/201014666},
      \href {https://ui.adsabs.harvard.edu/abs/2010A&A...518L.102A} {518, L102}
    
    \bibitem[\protect\citeauthoryear{{Arzoumanian} et~al.,}{{Arzoumanian}
      et~al.}{2011}]{2011A&A...529L...6A}
    {Arzoumanian} D.,  et~al., 2011, \mn@doi [\aap] {10.1051/0004-6361/201116596},
      \href {https://ui.adsabs.harvard.edu/abs/2011A&A...529L...6A} {529, L6}
    
    \bibitem[\protect\citeauthoryear{{Barreto-Mota}, {\VAN{Gouveia Dal
      Pino}{De}{de}}~Gouveia Dal~Pino, {Burkhart}, {Melioli}, {Santos-Lima}  \&
      {Kadowaki}}{{Barreto-Mota} et~al.}{2021}]{2021MNRAS.503.5425B}
    {Barreto-Mota} L.,  {\VAN{Gouveia Dal Pino}{De}{de}}~Gouveia Dal~Pino E.~M.,
      {Burkhart} B.,  {Melioli} C.,  {Santos-Lima} R.,   {Kadowaki} L.~H.~S.,
      2021, \mn@doi [\mnras] {10.1093/mnras/stab798}, \href
      {https://ui.adsabs.harvard.edu/abs/2021MNRAS.503.5425B} {503, 5425}
    
    \bibitem[\protect\citeauthoryear{{Bastien}}{{Bastien}}{1983}]{1983A&A...119..109B}
    {Bastien} P.,  1983, \aap, \href
      {https://ui.adsabs.harvard.edu/abs/1983A&A...119..109B} {119, 109}
    
    \bibitem[\protect\citeauthoryear{{Bhadari}, {Dewangan}, {Pirogov}  \&
      {Ojha}}{{Bhadari} et~al.}{2020}]{2020ApJ...899..167B}
    {Bhadari} N.~K.,  {Dewangan} L.~K.,  {Pirogov} L.~E.,   {Ojha} D.~K.,  2020,
      \mn@doi [\apj] {10.3847/1538-4357/aba2c6}, \href
      {https://ui.adsabs.harvard.edu/abs/2020ApJ...899..167B} {899, 167}
    
    \bibitem[\protect\citeauthoryear{{Bolatto}, {Wolfire}  \& {Leroy}}{{Bolatto}
      et~al.}{2013}]{2013ARA&A..51..207B}
    {Bolatto} A.~D.,  {Wolfire} M.,   {Leroy} A.~K.,  2013, \mn@doi [\araa]
      {10.1146/annurev-astro-082812-140944}, \href
      {https://ui.adsabs.harvard.edu/abs/2013ARA&A..51..207B} {51, 207}
    
    \bibitem[\protect\citeauthoryear{{Burkert} \& {Hartmann}}{{Burkert} \&
      {Hartmann}}{2004}]{2004ApJ...616..288B}
    {Burkert} A.,  {Hartmann} L.,  2004, \mn@doi [\apj] {10.1086/424895}, \href
      {https://ui.adsabs.harvard.edu/abs/2004ApJ...616..288B} {616, 288}
    
    \bibitem[\protect\citeauthoryear{{Cambr{\'e}sy}}{{Cambr{\'e}sy}}{1999}]{1999A&A...345..965C}
    {Cambr{\'e}sy} L.,  1999, \mn@doi [\aap] {10.48550/arXiv.astro-ph/9903149},
      \href {https://ui.adsabs.harvard.edu/abs/1999A&A...345..965C} {345, 965}
    
    \bibitem[\protect\citeauthoryear{{Clark}, {Peek}  \& {Putman}}{{Clark}
      et~al.}{2014}]{2014ApJ...789...82C}
    {Clark} S.~E.,  {Peek} J.~E.~G.,   {Putman} M.~E.,  2014, \mn@doi [\apj]
      {10.1088/0004-637X/789/1/82}, \href
      {https://ui.adsabs.harvard.edu/abs/2014ApJ...789...82C} {789, 82}
    
    \bibitem[\protect\citeauthoryear{{Dobbs} \& {Wurster}}{{Dobbs} \&
      {Wurster}}{2021}]{2021MNRAS.502.2285D}
    {Dobbs} C.~L.,  {Wurster} J.,  2021, \mn@doi [\mnras] {10.1093/mnras/stab150},
      \href {https://ui.adsabs.harvard.edu/abs/2021MNRAS.502.2285D} {502, 2285}
    
    \bibitem[\protect\citeauthoryear{{Duarte-Cabral} \& {Dobbs}}{{Duarte-Cabral} \&
      {Dobbs}}{2017}]{2017MNRAS.470.4261D}
    {Duarte-Cabral} A.,  {Dobbs} C.~L.,  2017, \mn@doi [\mnras]
      {10.1093/mnras/stx1524}, \href
      {https://ui.adsabs.harvard.edu/abs/2017MNRAS.470.4261D} {470, 4261}
    
    \bibitem[\protect\citeauthoryear{{Ester}, {Kriegel}, {Sander}  \& {Xu}}{{Ester}
      et~al.}{1996}]{1996kddm.conf..226E}
    {Ester} M.,  {Kriegel} H.-P.,  {Sander} J.,   {Xu} X.,  1996, in Second
      International Conference on Knowledge Discovery and Data Mining (KDD'96).
      Proceedings of a conference held August 2-4. pp 226--331
    
    \bibitem[\protect\citeauthoryear{{Falgarone}, {Pety}  \&
      {Phillips}}{{Falgarone} et~al.}{2001}]{2001ApJ...555..178F}
    {Falgarone} E.,  {Pety} J.,   {Phillips} T.~G.,  2001, \mn@doi [\apj]
      {10.1086/321483}, \href
      {https://ui.adsabs.harvard.edu/abs/2001ApJ...555..178F} {555, 178}
    
    \bibitem[\protect\citeauthoryear{{Federrath}}{{Federrath}}{2016}]{2016MNRAS.457..375F}
    {Federrath} C.,  2016, \mn@doi [\mnras] {10.1093/mnras/stv2880}, \href
      {https://ui.adsabs.harvard.edu/abs/2016MNRAS.457..375F} {457, 375}
    
    \bibitem[\protect\citeauthoryear{{Goodman} et~al.,}{{Goodman}
      et~al.}{2014}]{2014ApJ...797...53G}
    {Goodman} A.~A.,  et~al., 2014, \mn@doi [\apj] {10.1088/0004-637X/797/1/53},
      \href {https://ui.adsabs.harvard.edu/abs/2014ApJ...797...53G} {797, 53}
    
    \bibitem[\protect\citeauthoryear{{Hacar}, {Tafalla}, {Kauffmann}  \&
      {Kov{\'a}cs}}{{Hacar} et~al.}{2013}]{2013A&A...554A..55H}
    {Hacar} A.,  {Tafalla} M.,  {Kauffmann} J.,   {Kov{\'a}cs} A.,  2013, \mn@doi
      [\aap] {10.1051/0004-6361/201220090}, \href
      {https://ui.adsabs.harvard.edu/abs/2013A&A...554A..55H} {554, A55}
    
    \bibitem[\protect\citeauthoryear{{Hacar}, {Tafalla}, {Forbrich}, {Alves},
      {Meingast}, {Grossschedl}  \& {Teixeira}}{{Hacar}
      et~al.}{2018}]{2018A&A...610A..77H}
    {Hacar} A.,  {Tafalla} M.,  {Forbrich} J.,  {Alves} J.,  {Meingast} S.,
      {Grossschedl} J.,   {Teixeira} P.~S.,  2018, \mn@doi [\aap]
      {10.1051/0004-6361/201731894}, \href
      {https://ui.adsabs.harvard.edu/abs/2018A&A...610A..77H} {610, A77}
    
    \bibitem[\protect\citeauthoryear{{Hacar}, {Clark}, {Heitsch}, {Kainulainen},
      {Panopoulou}, {Seifried}  \& {Smith}}{{Hacar}
      et~al.}{2022}]{2022arXiv220309562H}
    {Hacar} A.,  {Clark} S.,  {Heitsch} F.,  {Kainulainen} J.,  {Panopoulou} G.,
      {Seifried} D.,   {Smith} R.,  2022, arXiv e-prints, \href
      {https://ui.adsabs.harvard.edu/abs/2022arXiv220309562H} {p. arXiv:2203.09562}
    
    \bibitem[\protect\citeauthoryear{{Heyer}, {Soler}  \& {Burkhart}}{{Heyer}
      et~al.}{2020}]{2020MNRAS.496.4546H}
    {Heyer} M.,  {Soler} J.~D.,   {Burkhart} B.,  2020, \mn@doi [\mnras]
      {10.1093/mnras/staa1760}, \href
      {https://ui.adsabs.harvard.edu/abs/2020MNRAS.496.4546H} {496, 4546}
    
    \bibitem[\protect\citeauthoryear{{Hily-Blant} \& {Falgarone}}{{Hily-Blant} \&
      {Falgarone}}{2009}]{2009A&A...500L..29H}
    {Hily-Blant} P.,  {Falgarone} E.,  2009, \mn@doi [\aap]
      {10.1051/0004-6361/200912296}, \href
      {https://ui.adsabs.harvard.edu/abs/2009A&A...500L..29H} {500, L29}
    
    \bibitem[\protect\citeauthoryear{{Inutsuka}, {Inoue}, {Iwasaki}  \&
      {Hosokawa}}{{Inutsuka} et~al.}{2015}]{2015A&A...580A..49I}
    {Inutsuka} S.-i.,  {Inoue} T.,  {Iwasaki} K.,   {Hosokawa} T.,  2015, \mn@doi
      [\aap] {10.1051/0004-6361/201425584}, \href
      {https://ui.adsabs.harvard.edu/abs/2015A&A...580A..49I} {580, A49}
    
    \bibitem[\protect\citeauthoryear{{Jackson}, {Finn}, {Chambers}, {Rathborne}  \&
      {Simon}}{{Jackson} et~al.}{2010}]{2010ApJ...719L.185J}
    {Jackson} J.~M.,  {Finn} S.~C.,  {Chambers} E.~T.,  {Rathborne} J.~M.,
      {Simon} R.,  2010, \mn@doi [\apjl] {10.1088/2041-8205/719/2/L185}, \href
      {https://ui.adsabs.harvard.edu/abs/2010ApJ...719L.185J} {719, L185}
    
    \bibitem[\protect\citeauthoryear{{K{\"o}nyves} et~al.,}{{K{\"o}nyves}
      et~al.}{2015}]{2015A&A...584A..91K}
    {K{\"o}nyves} V.,  et~al., 2015, \mn@doi [\aap] {10.1051/0004-6361/201525861},
      \href {https://ui.adsabs.harvard.edu/abs/2015A&A...584A..91K} {584, A91}
    
    \bibitem[\protect\citeauthoryear{{Li}, {Fang}, {Henning}  \&
      {Kainulainen}}{{Li} et~al.}{2013a}]{2013MNRAS.436.3707L}
    {Li} H.-b.,  {Fang} M.,  {Henning} T.,   {Kainulainen} J.,  2013a, \mn@doi
      [\mnras] {10.1093/mnras/stt1849}, \href
      {https://ui.adsabs.harvard.edu/abs/2013MNRAS.436.3707L} {436, 3707}
    
    \bibitem[\protect\citeauthoryear{{Li}, {Wyrowski}, {Menten}  \&
      {Belloche}}{{Li} et~al.}{2013b}]{2013A&A...559A..34L}
    {Li} G.-X.,  {Wyrowski} F.,  {Menten} K.,   {Belloche} A.,  2013b, \mn@doi
      [\aap] {10.1051/0004-6361/201322411}, \href
      {https://ui.adsabs.harvard.edu/abs/2013A&A...559A..34L} {559, A34}
    
    \bibitem[\protect\citeauthoryear{{McClure-Griffiths}, {Dickey}, {Gaensler},
      {Green}  \& {Haverkorn}}{{McClure-Griffiths}
      et~al.}{2006}]{2006ApJ...652.1339M}
    {McClure-Griffiths} N.~M.,  {Dickey} J.~M.,  {Gaensler} B.~M.,  {Green} A.~J.,
       {Haverkorn} M.,  2006, \mn@doi [\apj] {10.1086/508706}, \href
      {https://ui.adsabs.harvard.edu/abs/2006ApJ...652.1339M} {652, 1339}
    
    \bibitem[\protect\citeauthoryear{{Molinari} et~al.,}{{Molinari}
      et~al.}{2010}]{2010A&A...518L.100M}
    {Molinari} S.,  et~al., 2010, \mn@doi [\aap] {10.1051/0004-6361/201014659},
      \href {https://ui.adsabs.harvard.edu/abs/2010A&A...518L.100M} {518, L100}
    
    \bibitem[\protect\citeauthoryear{{Myers}}{{Myers}}{2009}]{2009ApJ...700.1609M}
    {Myers} P.~C.,  2009, \mn@doi [\apj] {10.1088/0004-637X/700/2/1609}, \href
      {https://ui.adsabs.harvard.edu/abs/2009ApJ...700.1609M} {700, 1609}
    
    \bibitem[\protect\citeauthoryear{{Ntormousi}, {Burkert}, {Fierlinger}  \&
      {Heitsch}}{{Ntormousi} et~al.}{2011}]{2011ApJ...731...13N}
    {Ntormousi} E.,  {Burkert} A.,  {Fierlinger} K.,   {Heitsch} F.,  2011, \mn@doi
      [\apj] {10.1088/0004-637X/731/1/13}, \href
      {https://ui.adsabs.harvard.edu/abs/2011ApJ...731...13N} {731, 13}
    
    \bibitem[\protect\citeauthoryear{{Palmeirim} et~al.,}{{Palmeirim}
      et~al.}{2013}]{2013A&A...550A..38P}
    {Palmeirim} P.,  et~al., 2013, \mn@doi [\aap] {10.1051/0004-6361/201220500},
      \href {https://ui.adsabs.harvard.edu/abs/2013A&A...550A..38P} {550, A38}
    
    \bibitem[\protect\citeauthoryear{{Peretto} et~al.,}{{Peretto}
      et~al.}{2012}]{2012A&A...541A..63P}
    {Peretto} N.,  et~al., 2012, \mn@doi [\aap] {10.1051/0004-6361/201118663},
      \href {https://ui.adsabs.harvard.edu/abs/2012A&A...541A..63P} {541, A63}
    
    \bibitem[\protect\citeauthoryear{{Pilbratt} et~al.,}{{Pilbratt}
      et~al.}{2010}]{2010A&A...518L...1P}
    {Pilbratt} G.~L.,  et~al., 2010, \mn@doi [\aap] {10.1051/0004-6361/201014759},
      \href {https://ui.adsabs.harvard.edu/abs/2010A&A...518L...1P} {518, L1}
    
    \bibitem[\protect\citeauthoryear{{Pineda} et~al.,}{{Pineda}
      et~al.}{2022}]{2022arXiv220503935P}
    {Pineda} J.~E.,  et~al., 2022, arXiv e-prints, \href
      {https://ui.adsabs.harvard.edu/abs/2022arXiv220503935P} {p. arXiv:2205.03935}
    
    \bibitem[\protect\citeauthoryear{{Planck Collaboration} et~al.,}{{Planck
      Collaboration} et~al.}{2016a}]{2016A&A...586A.135P}
    {Planck Collaboration} et~al., 2016a, \mn@doi [\aap]
      {10.1051/0004-6361/201425044}, \href
      {https://ui.adsabs.harvard.edu/abs/2016A&A...586A.135P} {586, A135}
    
    \bibitem[\protect\citeauthoryear{{Planck Collaboration} et~al.,}{{Planck
      Collaboration} et~al.}{2016b}]{2016A&A...586A.138P}
    {Planck Collaboration} et~al., 2016b, \mn@doi [\aap]
      {10.1051/0004-6361/201525896}, \href
      {https://ui.adsabs.harvard.edu/abs/2016A&A...586A.138P} {586, A138}
    
    \bibitem[\protect\citeauthoryear{{Ragan}, {Henning}, {Tackenberg}, {Beuther},
      {Johnston}, {Kainulainen}  \& {Linz}}{{Ragan}
      et~al.}{2014}]{2014A&A...568A..73R}
    {Ragan} S.~E.,  {Henning} T.,  {Tackenberg} J.,  {Beuther} H.,  {Johnston}
      K.~G.,  {Kainulainen} J.,   {Linz} H.,  2014, \mn@doi [\aap]
      {10.1051/0004-6361/201423401}, \href
      {https://ui.adsabs.harvard.edu/abs/2014A&A...568A..73R} {568, A73}
    
    \bibitem[\protect\citeauthoryear{{Reid}, {Dame}, {Menten}  \&
      {Brunthaler}}{{Reid} et~al.}{2016}]{2016ApJ...823...77R}
    {Reid} M.~J.,  {Dame} T.~M.,  {Menten} K.~M.,   {Brunthaler} A.,  2016, \mn@doi
      [\apj] {10.3847/0004-637X/823/2/77}, \href
      {https://ui.adsabs.harvard.edu/abs/2016ApJ...823...77R} {823, 77}
    
    \bibitem[\protect\citeauthoryear{{Soler}, {Hennebelle}, {Martin},
      {Miville-Desch{\^e}nes}, {Netterfield}  \& {Fissel}}{{Soler}
      et~al.}{2013}]{2013ApJ...774..128S}
    {Soler} J.~D.,  {Hennebelle} P.,  {Martin} P.~G.,  {Miville-Desch{\^e}nes}
      M.~A.,  {Netterfield} C.~B.,   {Fissel} L.~M.,  2013, \mn@doi [\apj]
      {10.1088/0004-637X/774/2/128}, \href
      {https://ui.adsabs.harvard.edu/abs/2013ApJ...774..128S} {774, 128}
    
    \bibitem[\protect\citeauthoryear{{Soler} et~al.,}{{Soler}
      et~al.}{2021}]{2021A&A...651L...4S}
    {Soler} J.~D.,  et~al., 2021, \mn@doi [\aap] {10.1051/0004-6361/202141327},
      \href {https://ui.adsabs.harvard.edu/abs/2021A&A...651L...4S} {651, L4}
    
    \bibitem[\protect\citeauthoryear{{Su} et~al.,}{{Su}
      et~al.}{2019}]{2019ApJS..240....9S}
    {Su} Y.,  et~al., 2019, \mn@doi [\apjs] {10.3847/1538-4365/aaf1c8}, \href
      {https://ui.adsabs.harvard.edu/abs/2019ApJS..240....9S} {240, 9}
    
    \bibitem[\protect\citeauthoryear{{Sun}, {Lu}, {Yang}, {Su}, {Zhang}, {Zhou}  \&
      {Lin}}{{Sun} et~al.}{2018}]{2018AcASn..59....3S}
    {Sun} J.~X.,  {Lu} D.~R.,  {Yang} J.,  {Su} Y.,  {Zhang} S.~B.,  {Zhou} X.,
      {Lin} Z.~H.,  2018, Acta Astronomica Sinica, \href
      {https://ui.adsabs.harvard.edu/abs/2018AcASn..59....3S} {59, 3}
    
    \bibitem[\protect\citeauthoryear{{Wang}, {Testi}, {Burkert}, {Walmsley},
      {Beuther}  \& {Henning}}{{Wang} et~al.}{2016}]{2016ApJS..226....9W}
    {Wang} K.,  {Testi} L.,  {Burkert} A.,  {Walmsley} C.~M.,  {Beuther} H.,
      {Henning} T.,  2016, \mn@doi [\apjs] {10.3847/0067-0049/226/1/9}, \href
      {https://ui.adsabs.harvard.edu/abs/2016ApJS..226....9W} {226, 9}
    
    \bibitem[\protect\citeauthoryear{{Yan}, {Yang}, {Sun}, {Su}, {Xu}, {Wang},
      {Zhou}  \& {Wang}}{{Yan} et~al.}{2021}]{2021A&A...645A.129Y}
    {Yan} Q.-Z.,  {Yang} J.,  {Sun} Y.,  {Su} Y.,  {Xu} Y.,  {Wang} H.,  {Zhou} X.,
        {Wang} C.,  2021, \mn@doi [\aap] {10.1051/0004-6361/202039768}, \href
      {https://ui.adsabs.harvard.edu/abs/2021A&A...645A.129Y} {645, A129}
    
    \bibitem[\protect\citeauthoryear{{Yuan} et~al.,}{{Yuan}
      et~al.}{2020}]{2020A&A...637A..67Y}
    {Yuan} L.,  et~al., 2020, \mn@doi [\aap] {10.1051/0004-6361/201936625}, \href
      {https://ui.adsabs.harvard.edu/abs/2020A&A...637A..67Y} {637, A67}
    
    \bibitem[\protect\citeauthoryear{{Yuan} et~al.,}{{Yuan}
      et~al.}{2021}]{2021ApJS..257...51Y}
    {Yuan} L.,  et~al., 2021, \mn@doi [\apjs] {10.3847/1538-4365/ac242a}, \href
      {https://ui.adsabs.harvard.edu/abs/2021ApJS..257...51Y} {257, 51}
    
    \bibitem[\protect\citeauthoryear{{Zernickel}, {Schilke}  \&
      {Smith}}{{Zernickel} et~al.}{2013}]{2013A&A...554L...2Z}
    {Zernickel} A.,  {Schilke} P.,   {Smith} R.~J.,  2013, \mn@doi [\aap]
      {10.1051/0004-6361/201321425}, \href
      {https://ui.adsabs.harvard.edu/abs/2013A&A...554L...2Z} {554, L2}
    
    \bibitem[\protect\citeauthoryear{{Zucker}, {Battersby}  \& {Goodman}}{{Zucker}
      et~al.}{2015}]{2015ApJ...815...23Z}
    {Zucker} C.,  {Battersby} C.,   {Goodman} A.,  2015, \mn@doi [\apj]
      {10.1088/0004-637X/815/1/23}, \href
      {https://ui.adsabs.harvard.edu/abs/2015ApJ...815...23Z} {815, 23}
    
    \makeatother
    \end{thebibliography}





\appendix
    

\section{Test of the method of removal of curved clouds}
\label{Test of the method of removal of curved clouds}
\begin{figure}
    \centering
    \includegraphics[width=\linewidth]{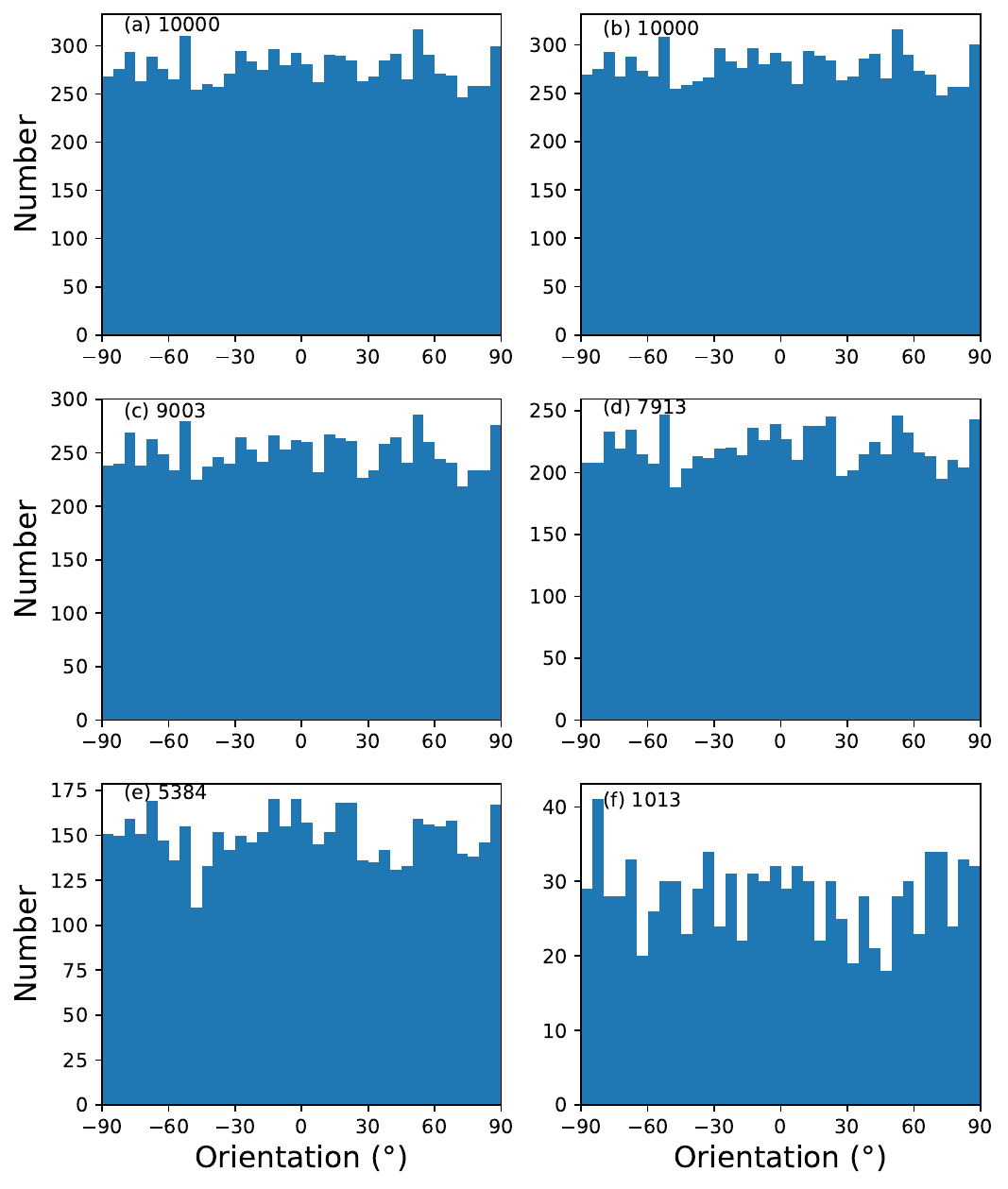}
    \caption{Orientation distribution for samples of mock filamentary clouds with different threshold for the $s$ parameter.  Panel (a): orientation distribution of the full 10000 mock filaments; here the histogram is based on the given value of the orientation angles. Panel (b): orientation of these filaments calculated using the elliptical fitting method. Panels (c, d, e, f): orientation distribution of subsamples obtained with different thresholds of the $s$ parameter for the removal of curved clouds (thresholds are 0.09, 0.07, 0.05, 0.03, respectively).  The numbers in each panel are the size of each subsample.}
\label{fig_a1}
\end{figure}

To ensure our sample selection is unbiased, here we test whether the method used in Section \ref{Removal of curved clouds} for removal of curved clouds would artificially introduce any pattern to the orientation distribution.  We generate 10,000 ellipses with varying aspect ratios and uniformly distributed orientations (from $-90^\circ$ to $90^\circ$).  Then we calculate their $s$ parameter using equation \ref{parameter s}.  It is expected that this $s$ parameter is related with the aspect ratio, but not the orientation.  We then apply different thresholds of $s$ to select different sets of subsample to see whether a preference for certain orientation angle would emerge. The results, depicted in Fig. (\ref{fig_a1}), indicate that the method introduces no orientation preference, thus we conclude that no bias is introduced during sample construction.

\section{Parameters of filaments we used}
\label{Parameters of filaments we used}
For the 522 filaments used for the analysis in this work, we list their relevant parameters in Table \ref{Table B1}. The complete table is available online. These parameters include the source name, orientation, coordinates, velocity, estimated distance, physical scale, velocity width, mass, maximum column density, and average column density.

    \begin{table*}
    \rotatebox{90}{%
    \begin{minipage}{\textheight}
    \centering
    \caption{The parameters of the sources we select in our sample. The complete table is available online.}
    \label{Table B1}
    \begin{tabular}{ccccccccccccc}
    \hline
    \hline
    Name & Orientation & l$\rm_{cen}$ & b$\rm_{cen}$ & V$\rm_{LSR}$ & Distance & Length & Width & Velocity width & Mass &$N(\text{H}_2)_\text{max}$ & $N(\text{H}_2)_\text{mean}$ & Parameter $p$\\
    & (deg) & (deg) & (deg) & (km s$^{-1}$) & (kpc) & (pc) & (pc) & (km s$^{-1}$) & M$_\odot$ &$10^{21}$cm$^{-2}$ &$10^{20}$cm$^{-2}$\\
    \hline
    $\text{G}113.775+2.783-52.77$ &  25.20 & 113.775 & 2.783 &  $-52.77$ & $2.66^{+0.12}_{-0.12}$ & 8.69 & 2.97 & 1.51 & 59.67 & 4.43 & 3.27 & 0.429\\
    $\text{G}137.808+4.450-53.11$ &  61.51 & 137.808 &  4.450 &   $-53.11$ & $1.95^{+0.04}_{-0.04}$ & 4.96 & 1.09 & 3.36 & 4.42 & 0.82 & 0.65 & 0.310\\
    $\text{G}132.267-0.583-53.11$ &   $-20.05$ & 132.267 &  $-0.583$ &  $-53.11$ & $1.96^{+0.04}_{-0.04}$ & 3.71 & 1.25 & 1.75 & 4.96 & 1.29 & 2.02 & 0.131\\
    $\text{G}108.808-0.417-53.27$ &   72.40 & 108.808 & $-0.417$ & $-53.27$ & $2.89^{+0.30}_{-0.30}$ & 4.09 & 1.34 & 1.82 & 5.81 & 0.98 & 1.85 & 0.295\\
    $\text{G}122.192+1.442-53.44$ &   33.23 & 122.192 &  1.422 & $-53.44$ & $2.51^{+0.29}_{-0.29}$ & 3.22 & 1.07 & 1.72 & 2.66 & 0.79 & 1.16 & 0.309\\
    $\text{G}116.058+3.733-53.11$ & 76.20 & 116.058 &  3.733 & $-53.11$ & $2.63^{+0.50}_{-0.50}$ & 4.71 & 1.21 & 2.17 & 3.17 & 0.55 & 0.76 & 0.160\\
    $\text{G}119.092+0.783-52.60$ & $-11.48$ & 119.092 &  0.783 & $-52.60$ & $2.61^{+0.28}_{-0.28}$ & 4.89 & 1.60 & 2.37 & 8.48 & 1.68 & 2.57 & 0.570\\
    $\text{G}118.292-0.333-52.77$ & $-35.80$ & 118.292 &  $-0.333$ & $-52.77$ & $2.66^{+0.28}_{-0.28}$ & 6.45 & 2.02 & 1.74 & 14.45 & 1.86 & 1.73 & 0.598\\
    $\text{G}148.933+0.567-52.94$ & 66.26 & 148.933 &  0.567 & $-52.94$ & $4.83^{+0.53}_{-0.53}$ & 8.45 & 2.60 & 1.06 & 19.80 & 1.03 & 1.49 & 0.393\\
    $\text{G}122.058+0.633-52.77$ & $-37.06$ & 122.058 & 0.633 & $-52.77$ & $2.52^{+0.28}_{-0.28}$ & 8.41 & 1.84 & 1.43 & 17.34 & 1.35 & 0.99 & 0.447\\
    
    \ldots & \ldots & \ldots & \ldots & \ldots & \ldots & \ldots & \ldots & \ldots & \ldots & \ldots & \ldots & \ldots\\

    \hline
    \end{tabular}
    \end{minipage}%
    }
    \end{table*}

\section{Example images of filaments with varying degrees of asymmetry}
\label{example images of filaments}
\begin{figure*}
    \centering
    \includegraphics[width=2\columnwidth]{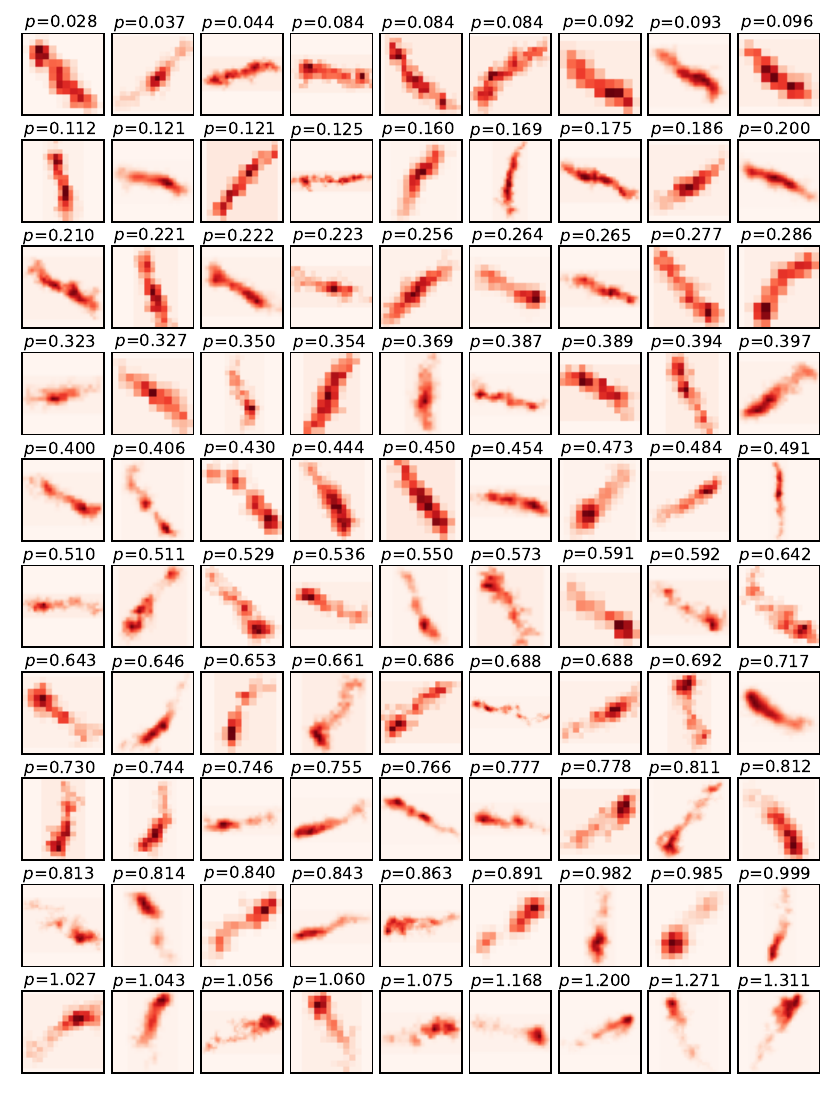}
    \caption{Example images of filaments with different value of parameter $p$.}
    \label{fig_c1}
\end{figure*}

In Fig. (\ref{fig_c1}), we present the $^{12}$CO $(J=1-0)$ velocity-integrated intensity maps of some filaments, sorted in ascending order of parameter $p$.

\section{Cross-check with different analysis method}
\label{Cross-check with different analysis method}
Since our analysis involves a sample selection step and thus cannot be directly applied to other dataset, we opt to cross-check our results using alternative methods. Specifically, we employ the orientation description method used by \citet{2013ApJ...774..128S}, to conduct a cross-check on our data.

This method primarily relies on the gradient of the density to characterize the directionality of the density structures. We employ Gaussian first derivative kernels in both horizontal and vertical directions to convolve the strength data separately, obtaining gradients in both directions. We use the Sobel operator (an approximate Gaussian derivative kernel) to perform this calculation:

\begin{equation}
\begin{aligned}
Sobel_x = \begin{pmatrix}
    -1 & 0 & 1 \\
    -2 & 0 & 2 \\
    -1 & 0 & 1 \\
\end{pmatrix}, \quad
Sobel_y = \begin{pmatrix}
    -1 & -2 & -1 \\
     0 &  0 &  0 \\
     1 &  2 &  1 \\
\end{pmatrix}.
\end{aligned}
\end{equation}

Then convolve the 2D integrated intensity with the Sobel operator:

\begin{equation}
    \frac{\partial I}{\partial x}=I * Sobel_x, \\
    \frac{\partial I}{\partial y}=I * Sobel_y, 
\end{equation}
where $I$ is the 2D integrated intensity.

According to equation (10) in \citet{2013ApJ...774..128S}, the `orientation description' of each pixel is given by:
\begin{equation}
    \psi = \arctan (\frac{\partial I/\partial x}{\partial I/\partial y}).
\end{equation}

\begin{figure}
    \centering
    \includegraphics[width=\linewidth]{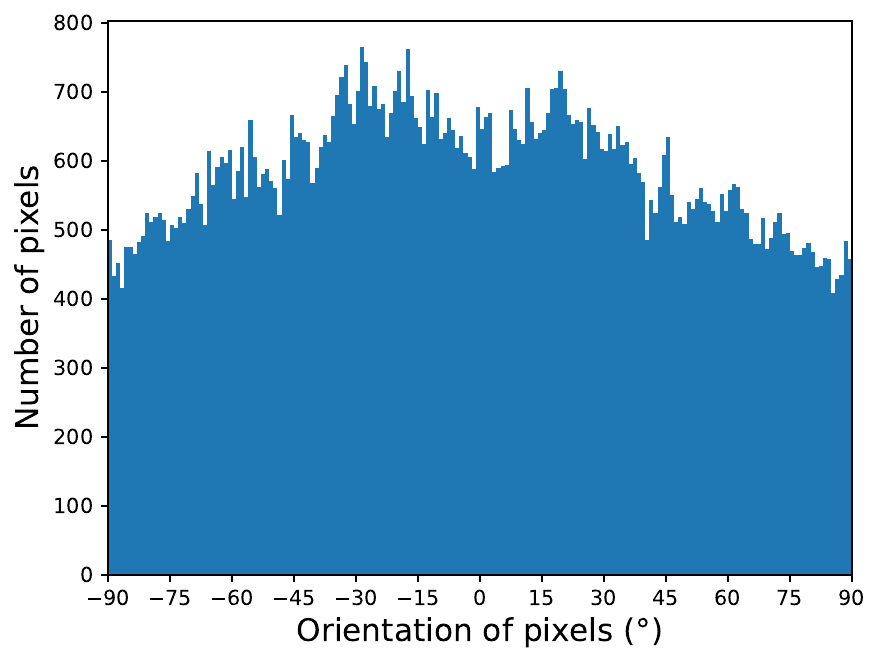}
    \caption{Distribution of pixel-based orientation in our sample, using the approach of \citet{2013ApJ...774..128S}.}
\label{fig_d1}
\end{figure}

Fig. (\ref{fig_d1}) shows the distribution of pixel-based orientations in the final samples calculated using the method proposed by Soler et al. We can see that the pixel-based orientation exhibits a weak bimodal distribution.

\begin{figure*}
    \centering
    \includegraphics[width=2\columnwidth]{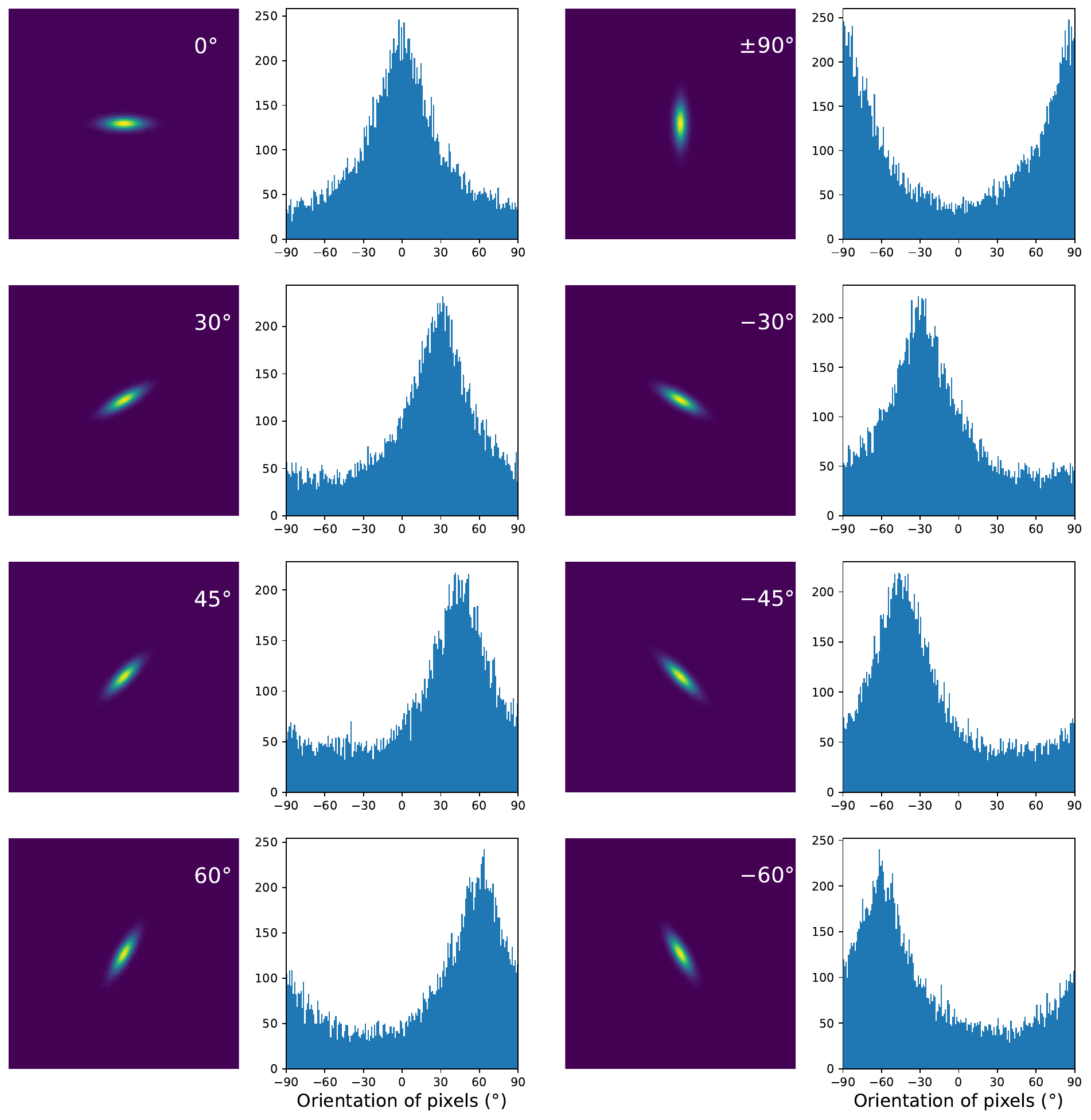}
    \caption{Pixel-based orientation distribution of individual Gaussian ellipse obtained with Soler's method, for a series of ellipses with well-defined orientation (labeled in each panel).}
\label{fig_d2}
\end{figure*}

The weakening of the bimodal feature could be related with that the Sobel operator is applied to all the pixels of a filament.  To test this, we generate a series of ellipses with well-defined orientations, while randomly setting the aspect ratio between 2 and 5. The intensity of these ellipses follows a two-dimensional Gaussian distribution overlaid with some Gaussian noise. Fig. (\ref{fig_d2}) displays the results for these ellipses, where we observe that the orientation of the pixels exhibits a certain degree of dispersion. This observation helps explain the dilution of the bimodal distribution seen in Fig. (\ref{fig_d1}).

\section{Orientation distribution of the tangents of a sinusoidal curve}
\label{secAppendixSinusoidal}

While we do not attempt to explain the origin of the bimodal distribution of the orientation of filaments (Figs. \ref{fig5}, \ref{fig6}) here, we do notice that a visual inspection of Fig. (\ref{fig9}) gives hints of some global sinusoidal patterns.
In the following we explore whether a bimodal distribution can be reproduced if the filaments are indeed organized as tangents to some (maybe more than one) sinusoidal patterns.

Given a curve specified by a function $y=f(x)$ in the $x${-}$y$ plane, its tangent at $x$ is given by $y' = f'(x)$, and the orientation angle is
$$
\theta = \arctan f'(x).
$$
For uniform sampling in the $x$ coordinates, the density distribution function of $\theta$ is
\begin{equation}
p(\theta) \propto \frac{dx}{d\theta} = \frac{1}{\cos^2\theta} \frac{1}{f''(x)} = \frac{1}{\cos^2\theta} \frac{1}{f''(f'^{-1}(\tan\theta))}.
\end{equation}

In the case of a sinusoidal curve, $f(x) = \sin(bx)$, we have
\begin{equation*}
\begin{split}
& f'(x) = b\cos(bx),\\
& f''(x) = -b^2\sin(bx),\\
& f'^{-1}(y) = \frac{1}{b}\arccos(\frac{y}{b}),
\end{split}
\end{equation*}
thus
\begin{equation}
p(\theta) \propto \frac{1}{\cos^2\theta} \frac{1}{\sqrt{b^2-\tan^2\theta}}.
\label{eqptheta}
\end{equation}
Obviously a phase shift of the sinusoidal curve (which may change $\sin$ into $\cos$, for example) does not change the shape of $p(\theta)$, and the coexistence of a few similar sinusoidal patterns will only widen the bimodal pattern.

\begin{figure}
\centering
\includegraphics[width=\linewidth]{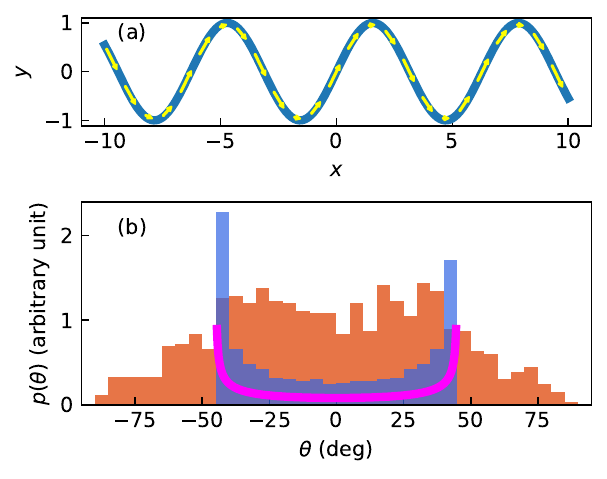}
\caption{Orientation distribution of the tangents of a sinusoidal curve.  Panel (a): an example sinusoidal curve.  The yellow arrows illustrate the filaments that are aligned with the curve.  Panel (b): distribution of the angle of tangential directions (relative to the $x$ axis).  The blue histogram is obtained from numerically differentiating the curve in panel (a) and then calculating the orientation angle of the tangents, while the orange histogram is obtained the same way excepts that noise is artificially added to the two components of the tangent vector.  The magenta curve is calculated using equation (\ref{eqptheta}).}
\label{fig_e1}
\end{figure}

Fig. (\ref{fig_e1}) shows an example for a sinusoidal curve, in which the density distribution of the tangent angles are calculated numerically (blue), numerically with added noise (orange), and analytically with equation (\ref{eqptheta}).

We can see that a bimodal distribution can be easily obtained, and if we are to fit the observed one (Fig. \ref{fig5}), adjusting the value of $b$ in $\sin(bx)$ and the added noise level would probably help.

Thus the conjectured existence of sinusoidal patterns may help explain the observed bimodal pattern, but that can't be the whole story.
It could be that, if the filaments are aligned with magnetic fields with more or less a sinusoidal configuration, for the location where the tangent angle is close to zero, the field strength is too low for the filaments to be well aligned.

\bsp	
\label{lastpage}
\end{document}